\documentclass[prd,aps,twocolumn,superscriptaddress,nofootinbib,showpacs,preprintnumbers,floatfix]{revtex4}
\usepackage{graphicx, epsfig, bm, amsmath}
\usepackage{amssymb}
\usepackage{xcolor}
\usepackage{float}
\definecolor{colorLink}{rgb}{0.7,0,0}
\definecolor{colorCite}{rgb}{0,.7,0}
\definecolor{colorURL}{rgb}{0,0,0.7}
\usepackage[colorlinks=true,linktocpage=true,linkcolor=blue,citecolor=colorCite,urlcolor=colorURL]{hyperref}

\usepackage{wasysym}
\usepackage{setspace}
\usepackage{multirow}

\newcommand{\DM}{{\scriptscriptstyle \text{DM}}}
\newcommand{\aW}{\alpha_{\scriptscriptstyle W}}
\newcommand{\mW}{m_{\scriptscriptstyle W}}

\newcommand{\thetaW}{\theta_{\scriptscriptstyle W}}

\definecolor{colorTC}{rgb}{.2,.7,.2}

\begin{document}

%%%%%%%%%%%%%%%%%%%%%%%%%%%%%%%%%%%%%%%%%%%

\preprint{IRFU-18-18}
\preprint{MIT-CTP 5041}
\preprint{LA-UR-18-28196}
%\preprint{\vbox{\vspace{30pt}\hbox{IRFU-18-18}}}
%\preprint{\vbox{\vspace{30pt}\hbox{MIT-CTP 5041}}}

\vspace*{15pt}
\title{Hunting for Heavy Winos in the Galactic Center}
\author{Lucia Rinchiuso}%\email{}
\affiliation{\!\!\mbox{ \footnotesize IRFU, CEA, D\'epartement de Physique des Particules, Universit{\'{e}} Paris-Saclay, F-91191 Gif-sur-Yvette, France}\,\,\,\vspace{0.7ex}}
\author{Nicholas L. Rodd}%\email{}
\affiliation{\footnotesize Center for Theoretical Physics, Massachusetts Institute of Technology, Cambridge, MA 02139, USA\vspace{0.7ex}}
\affiliation{\footnotesize Berkeley Center for Theoretical Physics, University of California, Berkeley, CA 94720, USA\vspace{0.7ex}}
\affiliation{\footnotesize Theoretical Physics Group, Lawrence Berkeley National Laboratory, Berkeley, CA 94720, USA\vspace{0.7ex}}
\author{Ian Moult}%\email{}
\affiliation{\footnotesize Berkeley Center for Theoretical Physics, University of California, Berkeley, CA 94720, USA\vspace{0.7ex}}
\affiliation{\footnotesize Theoretical Physics Group, Lawrence Berkeley National Laboratory, Berkeley, CA 94720, USA\vspace{0.7ex}}
\author{Emmanuel Moulin}%\email{} 
\affiliation{\!\!\mbox{ \footnotesize IRFU, CEA, D\'epartement de Physique des Particules, Universit{\'{e}} Paris-Saclay, F-91191 Gif-sur-Yvette, France}\,\,\,\vspace{0.7ex}}
\author{\\[2pt] Matthew Baumgart}%\email{}
\affiliation{\footnotesize Department of Physics, Arizona State University, Tempe, AZ 85287, USA\vspace{0.7ex}}
\author{Timothy Cohen}%\email{}
\affiliation{\footnotesize Institute of Theoretical Science, University of Oregon, Eugene, OR 97403, USA\vspace{0.7ex}}
\author{Tracy R. Slatyer}%\email{}
\affiliation{\footnotesize Center for Theoretical Physics, Massachusetts Institute of Technology, Cambridge, MA 02139, USA\vspace{0.7ex}}
\author{Iain W. Stewart}
\affiliation{\footnotesize Center for Theoretical Physics, Massachusetts Institute of Technology, Cambridge, MA 02139, USA\vspace{0.7ex}}
\author{Varun Vaidya}
\affiliation{\footnotesize Theoretical Division, MS B283, Los Alamos National Laboratory, Los Alamos, NM 87545, USA\vspace{0.7ex}}

\begin{abstract}
\begin{centering}
\vspace{5pt}
{\bf Abstract}\\[4pt]
\end{centering}
\noindent Observing gamma rays using ground-based atmospheric Cherenkov telescopes provides one of the only probes of heavy weakly interacting dark matter.  A canonical target is the thermal wino, for which the strongest limits come from searches for photon lines from annihilations in the Galactic Center. Irreducible finite energy resolution effects motivate refining the prediction for a wino signal beyond the photon line approximation; recently, modern effective field theory techniques have been utilized to obtain a precise calculation of the full photon energy spectrum from wino annihilation. In this paper, we investigate the implications for a realistic mock H.E.S.S.-like line search.  We emphasize the impact of including the non-trivial spectral shape, and we carefully treat the region of interest, presenting results for choices between $1^{\circ}$ and $4^{\circ}$ from the Galactic Center. Projected limits for wino masses from $1$-$70$ TeV are interpreted as a constraint on the wino annihilation rate, or alternatively as the minimum core size required such that the wino is not excluded.  If there is a thermal wino, H.E.S.S. will be able to probe cores of several kpc, which would begin to cause tension between this dark matter candidate and astrophysical observations/simulations.
\end{abstract}

\pacs{
95.35.+d, 95.85.Pw, 98.35.Jk, 98.35.Gi
}

\maketitle

\begin{spacing}{1.1}
\setcounter{section}{1}
Although dark matter (DM) constitutes more than 80\% of the matter in the universe, its nature remains elusive. While DM above the TeV scale would be difficult to produce at colliders, the relic particles can annihilate which would produce striking signals in gamma rays and cosmic rays (CRs). Imaging Atmospheric Cherenkov Telescopes (IACTs) have sensitivity to gamma rays from the hundred GeV to the hundred TeV scale, and can therefore provide an experimental handle on heavy DM annihilation, see \emph{e.g.}~\cite{Cahill-Rowley:2014boa,Lefranc:2016fgn}.

If such an annihilation signal exists, the highest photon statistics are likely to be found in the Galactic Center (GC) of the Milky Way, due to its proximity to the Earth, along with the expected accumulation of DM in the minimum of the Galactic gravitational potential well. However, the GC is a complicated region of the sky that produces photons at all wavelengths -- it contains regions of bright very-high-energy (VHE, $E\gtrsim 100 \text{ GeV}$) gamma-ray emission from conventional astrophysical processes, see \emph{e.g.}~\cite{vanEldik:2015qla,Moulin:2017cgb}.  Furthermore, IACTs have a substantial irreducible background due to the misidentification of CRs as gamma rays. Fortunately, the astrophysical backgrounds tend to have broad smooth spectra, implying that the cleanest and most convincing signal of DM annihilation in gamma rays would be a distinctive feature such as a spectral line.  Previous analyses~\cite{Abramowski:2013ax,Abdallah:2018qtu} have placed general model-independent limits on spectral lines using H.E.S.S. observations of the GC region.

In this paper, we focus on the wino, a prototypical DM candidate longtime for which these limits are very impactful~\cite{Cirelli:2008id,Fan:2013faa,Cohen:2013ama}.  The model is defined by extending the Standard Model by a single new electroweak triplet fermion with zero hypercharge, and the name \emph{wino} refers to the fact that this particle is identical to the superpartner of the electroweak bosons. Models including an electroweak triplet fermion as DM candidate were introduced from the 90's~\cite{Chardonnet:1993wd}. In this work, we explore the projected sensitivity of H.E.S.S. with the recently computed precision wino photon spectrum~\cite{Baumgart:2017nsr,us:NLL}, and using a mock H.E.S.S.-I-like observation of the GC. We show that the use of the full spectral shape leads to improved sensitivity to wino DM in a wide range of masses from the TeV to ten-TeV scale. Furthermore, a strategy relying on deep observations of the inner GC region would allow H.E.S.S. to explore thermally-produced winos for DM profiles with a core size up to several kpc.

The wino is a compelling target both due to the fact that it is arguably the simplest model of weakly interacting DM~\cite{Cirelli:2005uq}, and that it also could be the lightest superpartner, \emph{e.g.}~\cite{Giudice:1998xp,Randall:1998uk,Arvanitaki:2012ps,ArkaniHamed:2012gw}. Because its interactions are determined by the gauge structure of the Standard Model, the pure wino is highly predictive, since the DM mass is the only additional parameter relevant to phenomenology.  It is reasonable to treat the mass as a free parameter when interpreting limits from an experiment such as H.E.S.S.  However, one can require that the thermal relic abundance agrees with the measured value, implying a mass of $\sim 3$ TeV~\cite{Hisano:2006nn, Hryczuk:2010zi, Beneke:2016ync}. Maintaining the assumption of a thermal history, lower-mass winos can constitute a subdominant fraction of the DM, or a non-trivial cosmology can be invoked so that lighter winos could be all the DM.  Higher-mass winos are potentially viable DM candidates if their production and depletion mechanisms in the early universe differ from standard assumptions. Although searches at the LHC constrain the wino if its mass is below $\sim 450$ GeV \cite{Aaboud:2017mpt, Sirunyan:2018ldc}, TeV scale winos cannot be tested at the LHC~\cite{Low:2014cba,Cirelli:2014dsa,Gori:2014oua,Berlin:2015aba} or in direct detection experiments where their scattering cross section is near the neutrino floor \cite{Hill:2011be, Hill:2013hoa,Hisano:2015rsa}.  Finally, we note that measurements of the astrophysical anti-proton flux can also be used to constrain the wino~\cite{Hryczuk:2014hpa, Cuoco:2017iax}, although the associated systematic errors due to propagation uncertainty and the calculation of the spectrum can be large.

Recently, the first complete calculation of the full photon energy spectrum from heavy wino annihilation  based on modern effective field theory techniques has been presented~\cite{Baumgart:2017nsr,us:NLL}.  In particular, a number of effects that spoil the convergence of the standard fixed-order perturbative expansion can be treated with this formalism -- both the non-perturbative Sommerfeld enhancement and the resummation of large logarithms to next-to-leading-logarithmic accuracy are included.  This result provides the possibility to improve sensitivity to the thermal wino through the inclusion of the full photon spectrum, rather than only the gamma-ray line as has been the state-of-the-art until now.  Furthermore, a reliable estimate of the theoretical uncertainties can be extracted, and these are found to be $\sim 5\%$ for masses above $1$ TeV. Having a complete calculation of the photon spectrum allows experimental resolution effects to be properly incorporated, which is one of the goals of the work presented here.

Finally, it is worth emphasizing that the lessons learned here will be relevant to a variety of interesting DM candidates beyond the wino.  Another simple but challenging model to probe is the \emph{Higgsino}, where the Standard Model is extended by a pseudo-Dirac fermion that is an electroweak doublet with hypercharge 1/2.  The naming convention is again due to the charge assignment being the same as the superpartners of the two Higgs doublets required to supersymmetrize the Standard Model.  This model requires more structure than the wino, \emph{e.g.} some source of mass mixing is required to make the model safe from direct detection constraints.  A calculation of the thermal relic density points to a heavy mass scale -- the thermal Higgsino has a mass of $\sim 1 \text{ TeV}$.  This is another case where data from IACTs could be the dominant experimental probe, although other experiments are also relevant, for a recent discussion see~\cite{Krall:2017xij}.  Another set of important simple targets are minimal electroweak DM models, which have been cataloged in~\cite{Cirelli:2005uq,Cirelli:2009uv}.  The model building is motivated by simplicity -- these models are a classification of new particles that are only charged under the electroweak forces and can yield viable DM candidates.  In many cases, requiring a thermal history of the universe implies that the DM has a heavy mass, in some cases in the 10's of TeV range, such that IACTs are one of the dominant ways to test these theories.  We leave a characterization of the H.E.S.S. sensitivity in all these cases for future work, and here will focus exclusively on the wino.

An outline of this paper is as follows. Section~\ref{sec:model} discusses the ingredients required to model the signal and backgrounds, including a brief review of our calculation of the photon spectrum from wino annihilation. Section~\ref{sec:mockdata} defines the regions of interest relevant for our spatial analysis, describes the computation of the expected number of signal and background events, and discusses the statistical procedure used to derive our expected sensitivity. 
Section~\ref{sec:results} shows the results obtained on mock data of H.E.S.S.-like GC observations for various DM density core profiles, and the prospects with the current H.E.S.S. observation strategy of the GC region. The final section provides our conclusions.

%%%%%%%%%%%%%%%%%%%%%%%%%%%%%%%%%%%%%%%
\section{Hunting For Dark Matter}
\label{sec:model}
%%%%%%%%%%%%%%%%%%%%%%%%%%%%%%%%%%%%%%%

In this section, we review several prerequisites for understanding the search for DM annihilations from the GC.  We discuss the  DM density distribution with an emphasis on the variations that will be utilized to explore the dependence of our results on this uncertain quantity, the  relevant aspects of ground based observations with IACTs, and the recent work yielding the precision calculation of the photon spectrum used in the results that follow.

%%%%%%%%%%%%%%%%%%%%%%%%%%%%%%%%%%%%%%%
\subsection{Dark Matter Density Distribution}
%%%%%%%%%%%%%%%%%%%%%%%%%%%%%%%%%%%%%%%
The integrated photon flux due to the pair annihilation of DM particles from a region of solid angle $\Delta \Omega$ is computed using 
\begin{equation}
\frac{\text{d}\Phi^{\rm DM}_{\gamma}}{\text{d}E}\big(\Delta\Omega,E\big) = \frac{\langle\sigma\, v\rangle}{8\,\pi \, m_{\DM}^2}\frac{\text{d}N_{\gamma}(E)}{\text{d}E}\,J\big(\Delta\Omega\big) \,,
\label{eqn:flux}
\end{equation}
where $\langle\sigma\, v\rangle$ is the total annihilation cross section to all final states with a photon, $\text{d}N_\gamma/\text{d}E$ is the photon spectrum per such annihilation, $m_\DM$ is the DM mass, and $J(\Delta \Omega)$ is the integrated $J$-factor over a given region of interest (ROI) of solid angle size $\Delta \Omega$, defined by\footnote{We define the $J$-factor such that it carries units of [${\rm GeV}^2\cdot {\rm cm}^{-5}$]. Importantly, we do not associate the $\text{d} \Omega$ appearing in the definition with a unit of ${\rm sr}$, as this integral can be immediately identified as originating from a volume integral. Similarly, the $1/(4\, \pi)$ embedded in Eq.~\eqref{eqn:flux}, which originates in the surface area over which the flux from a given DM annihilation is diffused, is taken to be dimensionless.
A common alternative to this convention is to associate these quantities with a ${\rm sr}$ and ${\rm sr}^{-1}$ unit, respectively, as explained in detail in App.~A of~\cite{Lisanti:2017qoz}. Note in either convention the units for the flux $\Phi_{\gamma}^{\DM}$ is identical.
}
\begin{equation} 
J\big(\Delta \Omega\big) \equiv \int_{\Delta \Omega} \text{d}\Omega \int_0^\infty \text{d}s\, \rho_\DM\big(r(s,\theta)\big)^2 \,,
\label{eqn:jfac} 
\end{equation} 
where $\rho_\DM$ is the mass density of the DM.  We take the standard observer centered coordinate system so that $r =  \big(s^2 +r_{\odot}^2-2\,r_{\odot}\,s\, \cos\theta \big)^{1/2}$, where $s$ is the distance along the line of sight from the observer to the annihilation point, $r_{\odot} = 8.5 \text{ kpc}$ is the distance from the Sun to the GC, and $\theta$ is the angle between the direction of observation and the Galactic center.

If no significant excess over the background is found, indirect searches for DM annihilation signals can be interpreted as an upper limit on the annihilation cross section into a specific final state and for a given value of the $J$-factor, which parameterizes the DM density in the ROI. If the DM density in that region is not well-known, constraints should be interpreted as a joint limit on the particle physics and the astrophysics; for a particular particle physics model, constraints on the $J$-factor can be set, and if those constraints are inconsistent with the known constraints on the DM density distribution, we can then say that the model is ruled out.

The DM density in the region close to the GC has large uncertainties, because the density of visible baryonic matter is expected to dominate that of DM at small Galactocentric radii. On the observational front, this means that going from gravitational measurements of the total mass density to limits on the DM density requires careful modeling of the baryonic component and has associated large systematic uncertainties, see, for instance,~\cite{Iocco:2015xga,2017MNRAS.465.1621P}. Simulation-based predictions including hydrodynamics
and feedback physics in addition to the gravitational effects for the expected DM abundance have large uncertainties due to the effects of baryonic physics, and at sufficiently small Galactocentric distances, the resolution limit of simulations also becomes relevant. These issues currently prevent them from making robust predictions for the DM profile at radii smaller than a few kpc.

\begin{figure}[t] 
\begin{center}
\includegraphics[width=0.48\textwidth]{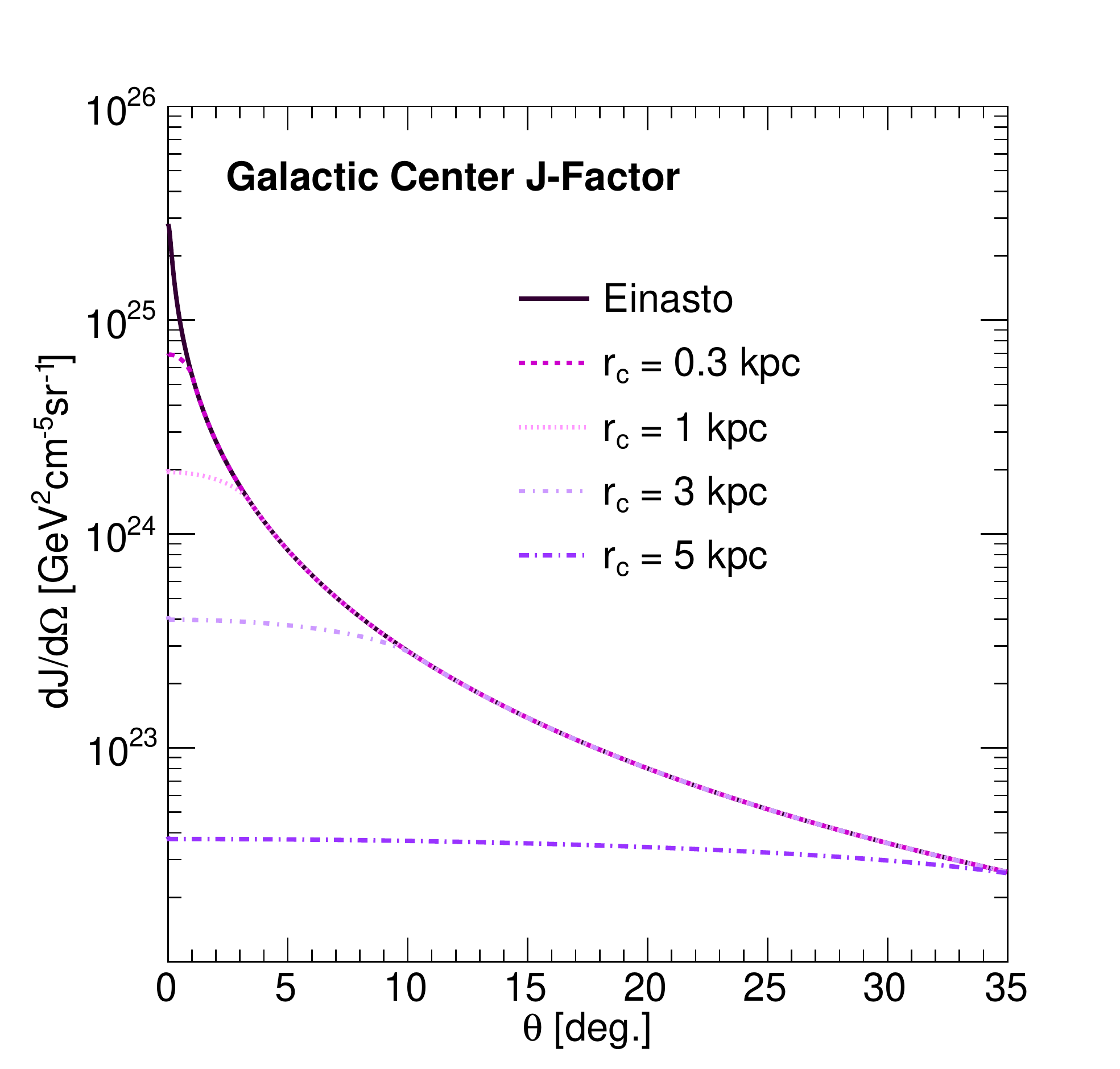}
\caption{$J$-factors 
for different DM profiles as a function of the angle $\theta$ from the GC. The DM profiles chosen in this study are the Einasto profile (solid line) and cored profiles with core radii of 0.3 kpc (dashed line), 1 kpc (dotted line), 3 kpc (dotted-dashed line), and 5 kpc (long-dashed-dotted line).}
\label{fig:profiles}
\end{center}
\end{figure} 

Simulations including only DM particles and neglecting the baryonic physics give rise to cusped density profiles that rise steeply toward the GC; such profiles are often parameterized by the Navarro-Frenk-White (NFW) \cite{Navarro:1995iw} or Einasto \cite{1965TrAlm...5...87E} profiles
\begin{equation} 
\rho_\DM(r) = \rho_0 \begin{cases} 
\Big[ \Big(\frac{r}{r_s}\Big) \Big(1 + \left(\frac{r}{r_s}\right )^2 \Big)\Big]^{-1} & \text{NFW}
\\
\\
\exp\left[-\frac{2}{\alpha} \left( \left(\frac{r}{r_s}\right)^\alpha - 1 \right)\right] & \text{Einasto}\end{cases} \,,
\end{equation}
where $r$ is the distance from the GC and $r_{\rm s}$ is a scale radius determined from the simulation. 
In this work we will use the Einasto profile as our baseline for a cusped DM profile, with the same parameters as in~\cite{Abramowski:2011hc}: explicitly, we choose $\alpha = 0.17$, $r_s=20$ kpc \cite{Pieri:2009je}, and $\rho_0$ chosen so that $\rho_\DM(r_{\odot}) = 0.39$ GeV/cm$^3$. This last choice is based on estimates of the DM density at the position of the Earth \cite{Catena:2009mf}.

Once baryonic matter is included in simulations, the short-distance cusp can be flattened out, producing a ``cored'' profile. For Milky-Way-sized galaxies, the scale within which the DM density is flattened can be of order 1 kpc~\cite{Chan:2015tna}. Depending on the modeling of baryonic physics within the simulation, DM cores in Milky Way-like galaxies extending to $\sim5$ kpc can be obtained~\cite{Mollitor:2014ara}.
On the observational front, the total DM mass in the Galactic Bulge region can be estimated from measurements of Bulge stellar populations \cite{2015MNRAS.448..713P}, and disfavors a NFW profile with a core size exceeding $\sim 2$ kpc \cite{Hooper:2016ggc}. However, a recent analysis using a dynamical modeling of the Galactic bulge, bar, and disk favors a shallow cusp or core in the Bulge region~\cite{2017MNRAS.465.1621P}. In order to account for possible kpc-sized DM cores in the GC, we will empirically parameterize a core of varying sizes by using the Einasto profile described above for $r > r_\text{\rm c}$, and setting $\rho_\DM(r) = \rho_\DM(r_\text{c}) = \rho_{\rm Einasto}(r_\text{\rm c})$ for $r < r_\text{\rm c}$. The normalization of the profile is such that $\rho_\DM(r_{\odot})$ = $\rho_\odot$. We plot the $J$-factor versus the angular distance, $\theta$, between the GC and the observation direction, for the Einasto profile and several choices of the core size, in Fig.~\ref{fig:profiles}.

%%%%%%%%%%%%%%%%%%%%%%%%%%%%%%%%%%%%%%%
\subsection{Ground-Based Observations with IACTs}
\label{sec:GCTeV}
%%%%%%%%%%%%%%%%%%%%%%%%%%%%%%%%%%%%%%%

The most promising avenue for experimental tests of wino DM lies in indirect detection; since the relevant mass scales are high, IACTs have sensitivity to the annihilation products of thermal winos. Furthermore, the cross section for high-mass weakly-interacting DM annihilation can be strongly enhanced at low velocities by the nonperturbative Sommerfeld enhancement \cite{Hisano:2003ec,Hisano:2004ds}, which also enhances the gamma-ray line signal relative to the continuum emission. The enhancement effect is large for the thermal wino, and so the pure wino DM presents a particularly attractive target for gamma-ray line searches with IACTs \cite{Fan:2013faa,Cohen:2013ama}.

Current arrays of IACTs like H.E.S.S., MAGIC, and VERITAS consist of 2-to-5 telescopes on the ground. The differential flux sensitivity achieved is $10^{-12}\text{ TeV}^{-1}\text{ cm}^{-2}\text { s}^{-1}$ at $\sim1\text{ TeV}$, about 1\% of the Crab flux~\cite{Aharonian:2006pe}. Based in Namibia near the tropic of Capricorn, the H.E.S.S. observatory is particularly well located to observe the central region of the Milky Way. Phase I of H.E.S.S. consists of four 12 m-diameter telescopes and reaches an angular resolution of 0.06$^{\circ}$ (68\% containment radius) and an energy resolution $\Delta E/E$ of 10\% above 300 GeV~\cite{2009APh32231D}. 

The GC region harbors numerous VHE gamma-ray emissions: they include H.E.S.S. J1745-290~\cite{Aharonian:2004wa,Aharonian:2009zk} a strong emission coincident with supermassive black hole Sagittarius A*, 
 the supernova/pulsar wind nebula G0.9+0.1~\cite{Aharonian:2005br}, the supernova remnant H.E.S.S. J1745-303~\cite{Aharonian:2008gw}, and a diffuse emission extending along the Galactic plane~\cite{Aharonian:2006au,HESS2014sla,Abramowski:2016mir}. 
The H.E.S.S. observatory has carried out a deep observation program of the GC region from 2004 to 2014. The rich observational dataset obtained from H.E.S.S. phase I has been used to look for continuum~\cite{Abramowski:2011hc,Abdallah:2016ygi} and line~\cite{Abramowski:2013ax,Abdallah:2018qtu} signals from DM annihilations. Standard analyses of H.E.S.S.-I observations of the GC region provided  $\sim 250$ hours of live time in the inner 1$^{\circ}$ of the GC with a mean zenith angle of about 20$^{\circ}$ that yields an energy threshold of 160 GeV.   The energy-dependent gamma-ray acceptance reaches $\sim3\times10^5\text{ m}^2$ above 1 TeV, with a typical hadronic rejection factor of about $10$. A rejection factor of $10$ corresponds to an efficiency of $10\%$, where the efficiency is defined by the number of events passing the overall event selection cuts.

In order to face the challenging standard astrophysical backgrounds, a robust approach consists of masking these regions from the data analysis for DM searches, as successfully applied in~\cite{Abdallah:2016ygi,Abdallah:2018qtu}. Once this has been performed, the dominant background in the GC region consists of misidentified CR hadrons (protons and nuclei), electrons,\footnote{The CR electron spatial distribution is assumed isotropic. No significant anisotropy of the VHE CR electrons is found in Fermi-LAT observations on any angular scale~\cite{Abdollahi:2017kyf}.} and Galactic diffuse emission. The dominant flux of CR hadrons interacting in the Earth's atmosphere generates
hadronic showers which include electromagnetic sub-showers from neutral pions decaying into photons. Hadronic showers can be efficiently discriminated from the shower initiated by primary gamma-rays, requiring a stereoscopic view of the event and using morphological and timing parameters of the shower image in the camera. The incoming CR hadron flux is much larger than the CR electrons and gamma-ray fluxes, so that a fraction of the hadron flux cannot be rejected due to the finite hadron rejection power of the instrument.

\begin{figure*}[t] 
\begin{center}
\raisebox{0.75cm}{
\includegraphics[width=0.45\textwidth]{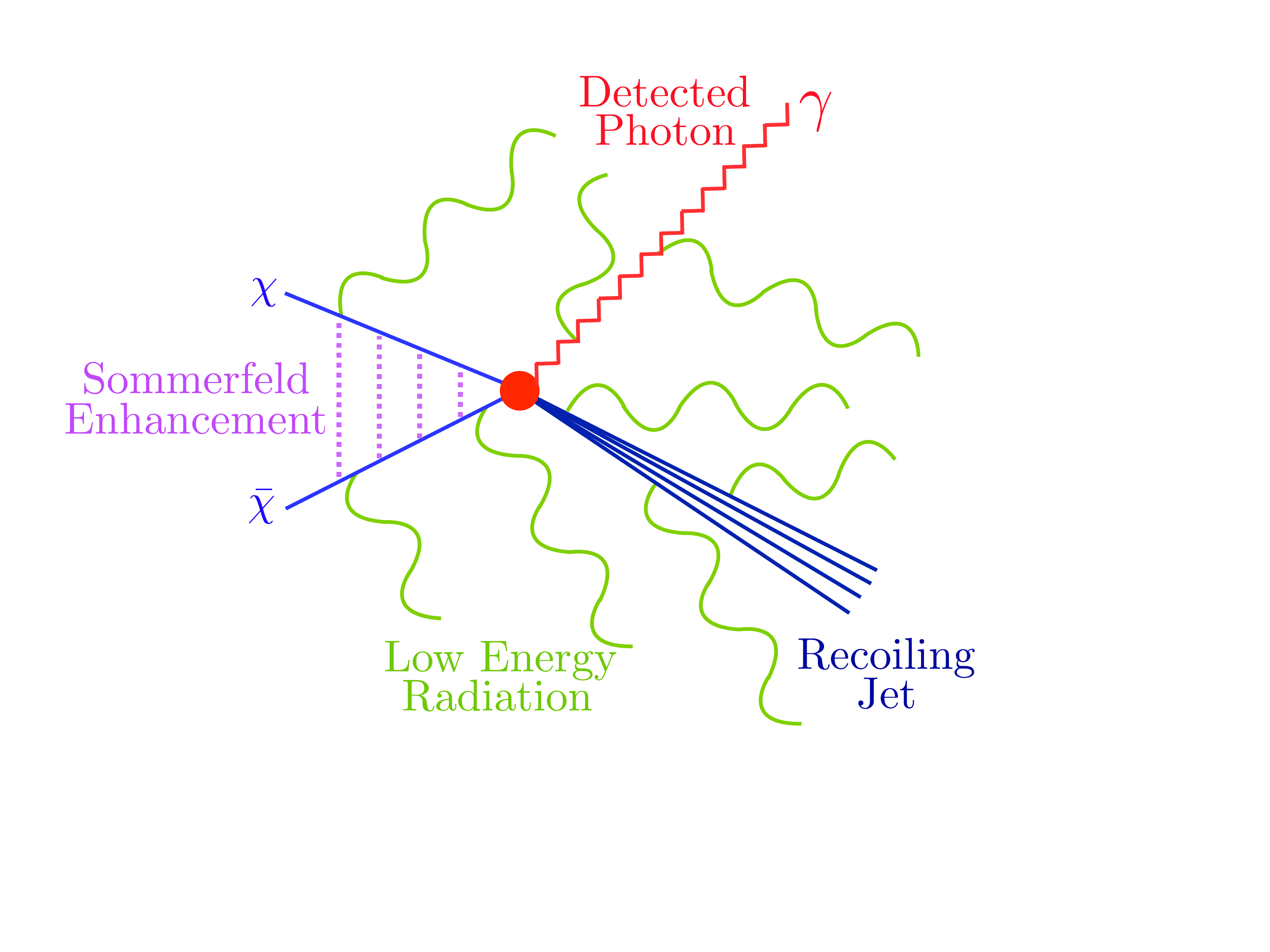}}
\hfill
\includegraphics[width=0.45\textwidth]{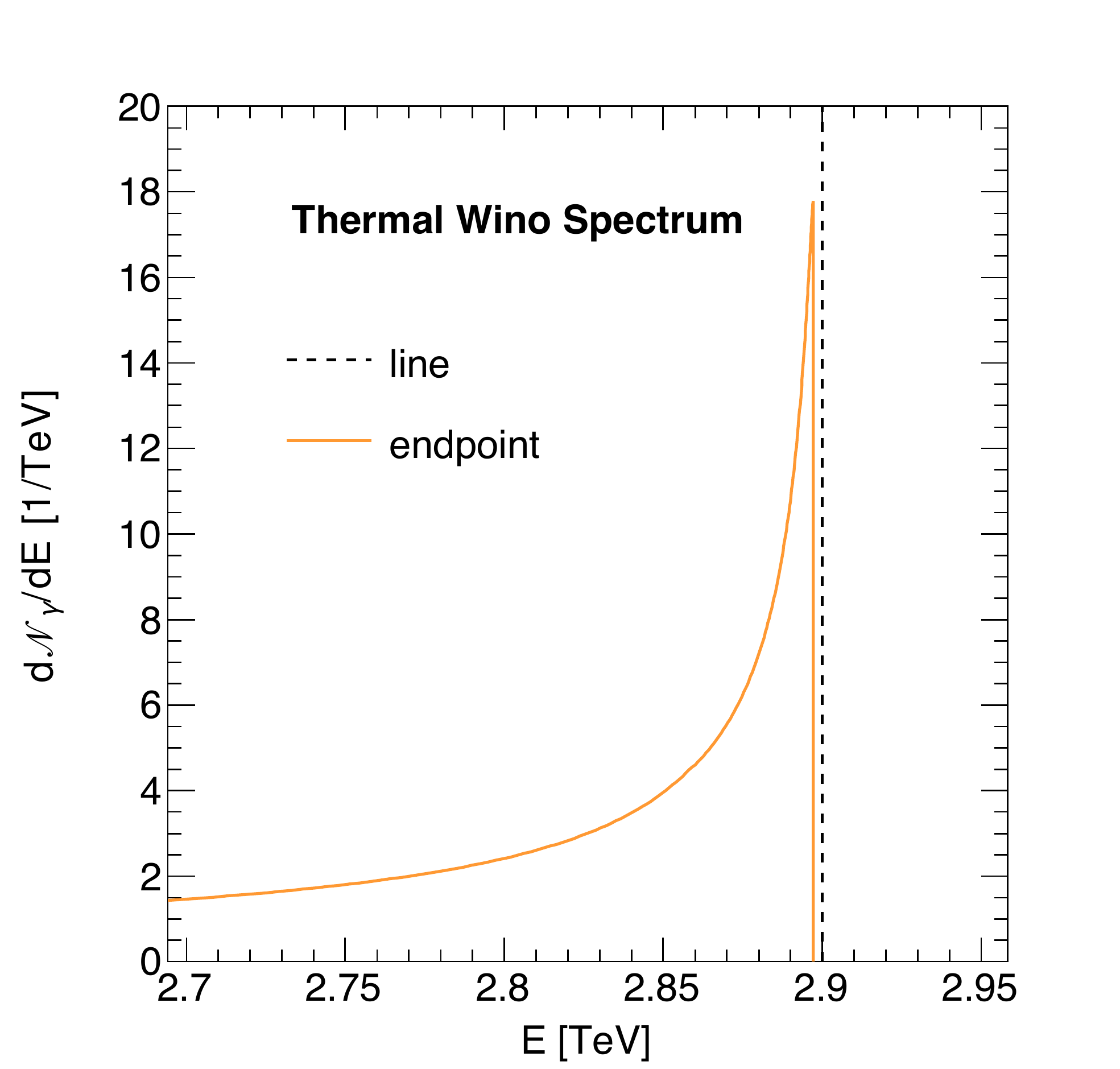}
\caption{{\it Left:} A schematic of the key physical contributions for the precision calculation as relevant for a H.E.S.S. search for heavy wino annihilation. DM particles, $\chi$, annihilate to the detected photon (in red), which recoils against a jet (in blue), \emph{i.e.} a collimated spray of electroweak radiation.  Low energy isotropic radiation (in green) also yields important physical effects. The winos collide with non-relativistic velocity, and the exchange of electroweak gauge bosons gives rise to the Sommerfeld enhancement (in purple).
{\it Right:} The spectral shape of the endpoint contribution at NLL (orange solid line), as compared to the line which is a pure delta function (black dashed line). The theoretical spectrum is shown for a thermal wino of mass $m_{\rm DM} = 2.9$ TeV.
}
\label{fig:theory_setup}
\end{center}
\end{figure*}

The measurement of the residual background in the GC region is complex~\cite{Abdallah:2016ygi,Abdallah:2018qtu}. An accurate background 
determination can be obtained for each observation where the background events are recorded in a region symmetric to the signal region from the pointing position. This allows the signal region and background regions to have the same sky acceptance and solid angle size, and thus does not require further offline normalization. This technique is very well suited in the case where a strong emission gradient is expected between the signal and background regions. However, the technique weakens for diffuse emission that is extended on the scale of the field of view of the instrument. In particular, for dark matter searches, this technique is proven to be very efficient in the case of cuspy DM density distributions, 
but fails for cored profiles with flattened density within 100 parsec or more of the GC. 
In order to avoid this limitation, one can extract the residual background energy distribution from extragalactic observations~\cite{Abdallah:2016ygi}. In this case, the residual background is extracted from blank fields at high Galactic latitudes in the same observation conditions as for the GC dataset.
Alternatively, Monte Carlo simulations can be used to predict the residual background rate, since they can be performed in the same observational conditions and telescope configurations as for the GC dataset, allowing for reduced systematic uncertainties~\cite{Holler:2017ynz}.

%%%%%%%%%%%%%%%%%%%%%%%%%%%%%%%%%%%%%%%
\subsection{Precision Signal Spectrum}
%%%%%%%%%%%%%%%%%%%%%%%%%%%%%%%%%%%%%%%

Due to the large backgrounds, the most striking signal for DM annihilation is a line signal, which in this energy range should not arise from astrophysical backgrounds. However, because of the finite energy resolution of IACTs, it is impossible to measure only the line spectrum; gamma-ray photons from DM annihilation to $\gamma\, \gamma$ or $\gamma\, Z$ will inevitably be accompanied by gamma-ray photons from other annihilation final states, and these cannot be distinguished on an event-by-event basis if the energy of the resulting photon varies by an amount less than the energy resolution of the telescope. Furthermore, if a smooth background model is included in the fit, as in \cite{Abramowski:2013ax}, the unaccounted-for presence of lower-energy signal photons could potentially bias the background model.  Thus to obtain precise and accurate constraints, it is important to have a theoretical prediction for the full photon spectrum to compare with the data, rather than simply comparing constraints on an isolated line to a theoretical prediction for the strength of the line signal.  

Obtaining a reliable prediction for the photon spectrum from wino annihilation is complicated by the presence of the hierarchical scales $\mW$, and $m_\DM$, and in the endpoint region -- where 
\begin{equation}
z\equiv \frac{E}{m_\DM}\,,
\label{eq:z}
\end{equation}
is close to 1 -- by the presence of non-trivial phase space restrictions. By looking for a line signal within the H.E.S.S. resolution of $m_\DM$, the final state must be close to a two-body decay, for which $z=1$. More precisely, it must consist of collimated energetic radiation recoiling against the detected photon (an electroweak jet), as well as additional low energy radiation. Any additional radiation would make the energy of the photon far from $m_\DM$. This configuration is shown on the left of Fig.~\ref{fig:theory_setup}. This restriction introduces perturbative Sudakov double logarithms $\aW\log^2(m_\DM/\mW)$~\cite{Hryczuk:2011vi,Baumgart:2014vma,Bauer:2014ula,Ovanesyan:2014fwa,Baumgart:2014saa,Baumgart:2015bpa,Ovanesyan:2016vkk}, and $\aW\log^2(1-z)$  \cite{Baumgart:2014vma,Baumgart:2014saa,Baumgart:2015bpa,Baumgart:2017nsr} , as well as Sommerfeld enhancement terms of the form $\left( \aW m_\DM/\mW \right )^k$ \cite{Hisano:2003ec,Hisano:2004ds,Cirelli:2007xd,ArkaniHamed:2008qn,Blum:2016nrz}. Here $\aW$ is the weak fine structure constant. Reliable predictions for the shape of the distribution in the endpoint region require that all these effects are resummed to all orders in perturbation theory. Once the perturbative series is reorganized in this manner, it again converges rapidly due to the smallness of the electroweak coupling, $\aW$, allowing precise theoretical predictions for the photon spectrum.

In \cite{Baumgart:2017nsr}, an effective field theory (EFT) framework was developed for the calculation of the photon spectrum in the endpoint region for heavy DM annihilation. It combines non-relativistic EFTs for the description of the annihilating DM, and soft-collinear effective theory (SCET) \cite{Bauer:2000yr, Bauer:2001ct, Bauer:2001yt}, as well as its multi-scale extensions \cite{Bauer:2011uc,Larkoski:2014tva,Procura:2014cba,Larkoski:2015zka,Pietrulewicz:2016nwo}, and extensions to include massive gauge bosons~\cite{Chiu:2007yn,Chiu:2008vv,Chiu:2007dg}, for the treatment of the final state radiation. This EFT allows the photon energy spectrum to be computed precisely, properly incorporating both the Sommerfeld and Sudakov effects (as well as their interplay) to all orders, and allowing for reliable uncertainty estimates.

Using this EFT, an analytic form for the photon energy spectrum in the endpoint region for annihilating pure wino DM was derived in~\cite{Baumgart:2017nsr} at leading logarithmic (LL) accuracy, and this calculation was extended in~\cite{us:NLL} to next-to-leading logarithmic (NLL) accuracy, greatly reducing the theoretical uncertainty. For simplicity, we present the final formula for the LL photon spectrum in this region, as this allows us to illustrate the general features of its shape in the endpoint region with a simple expression. We then briefly comment on how this is modified by additional logarithmic corrections at NLL.
We refer the reader to~\cite{Baumgart:2017nsr} for the derivation of the LL result, and \cite{us:NLL} for the analytic form of the spectrum at NLL.

In the endpoint region, the photon spectrum at LL accuracy can be written as a function of $z$ and $m_\DM$ as
\newpage
\begin{widetext}
\begin{align}
\left(\frac{\text{d} \sigma}{\text{d}z}\right)^{\text{LL}} \!&=\,
  4 \,|s_{0\pm}|^2\, \hat \sigma^\text{LL}_{\text{line}}\, \delta(1-z)   + \frac{2\,\aW}{\pi}   \frac{\hat \sigma^\text{LL}_{\text{line}}}{1-z} \,e^{\frac{4\,\aW}{\pi} \, L_J^2(z)}  \bigg\{  F_1  \Big( 3\,L_S(z) - 2\,L_J(z) \Big)   e^{\frac{-3\,\aW}{\pi}\, L^2_S(z)  }    -   2\,F_0\, L_J(z) \bigg\} \,.
\label{eq:resummed}
\end{align}
\end{widetext}
This provides a simple analytic expression describing both the line contribution, which is given by the first term in Eq.~(\ref{eq:resummed}) proportional to $\delta(1-z)$, as well as the endpoint contribution which is given by the second term, and is a non-trivial function of $z$, describing the steeply falling spectrum. The line contributions were first calculated with resummation in \cite{Bauer:2014ula,Ovanesyan:2014fwa,Ovanesyan:2016vkk}, while it is the shape of the spectrum away from $z=1$ that is the primary contribution of \cite{Baumgart:2017nsr}. On the right of Fig.~\ref{fig:theory_setup} we show the spectrum of photons associated with the endpoint for the thermal wino.

We now describe each of the components of Eq.~(\ref{eq:resummed}) in turn. Both terms are multiplied by the exclusive line cross section (without Sommerfeld effects), which at leading logarithmic accuracy is given by
\begin{align}
\hspace{-2pt}\hat \sigma^\text{LL}_{\text{line}}= {\pi \,\aW^2\, \sin^2 \thetaW \over 2\,m_\DM^2\, v} \exp\left[- \frac{4\,\aW}{\pi}\, \ln^2 \left( \frac{\mW}{2\,m_\DM}  \right) \right] \,,
\label{eq:partonicSigmaLine}
\end{align}
where $\thetaW$ is the Standard Model weak mixing angle. This can be computed within an EFT framework by considering charged wino annihilation into both $\gamma\,\gamma$ and $\gamma\,Z$.
The exponential appearing in this formula is the massive Sudakov form factor \cite{Collins:1989bt} and is due to the exchange of virtual electroweak bosons. 
This process is then mapped onto the neutral wino initial state by non-trivial mixing due to the Sommerfeld enhancement involving the exchange of a ladder of gauge bosons with one or more $W^\pm$ bosons, \emph{i.e.}, $s_{0\pm} \neq 0$.

The energy dependence of the photon spectrum in the endpoint region, which is crucial to our analysis, is described by a $1/(1-z)$ power law growth towards the endpoint, modified by the logarithms
\begin{align}
L_J(z)&= \ln \left({\mW/m_\DM \over 2 \, \sqrt{1-z} }\right) \Theta\!\left(1-\frac{\mW^2}{4\, m_\DM^2}-z \right)\,, \nonumber \\[5pt]
L_S(z)&= \ln \left({\mW/m_\DM \over 2 \, (1-z) }\right)  \Theta\!\left( 1-\frac{\mW}{2\,m_\DM}-z \right) \,,
\label{eq:LJLS}
\end{align}
associated with additional radiation in the final state. The $\Theta$ functions are set by the kinematics, and cut off the divergence in the $1/(1-z)$ growth before reaching the $z=1$ endpoint. Importantly, this power law form is unmodified beyond LL, with higher order corrections simply dressing this result with additional logarithms. Furthermore, we find that these higher order corrections are of the anticipated size, and that their primary utility is to reduce the theoretical uncertainty.

Finally, the non-perturbative Sommerfeld effect is captured by a non-relativistic quantum mechanics calculation of the matrix element of the  $S$-wave combination for the annihilating neutral winos $(\chi^0\chi^0)_S$,  
\begin{align} \label{eq:wavefunction}
   &\Big\langle 0 \Big|\, \chi^{0\,T}\, i\sigma_2 \,\chi^{0\,\,} \,\Big| \big(\chi^0 \chi^0\big)_S \Big\rangle = 4 \sqrt{2} \, m_\DM\, s_{00} \,, \nonumber \\[5pt]
  &\Big\langle 0 \Big|\, \chi^{+T}\, i\sigma_2 \,\chi^-\, \Big| \big(\chi^0 \chi^0\big)_S \Big\rangle= 4\, m_\DM \,s_{0\pm} \,,
\end{align}
 where $\sigma_2$ is the second Pauli matrix, and we have used the standard notation $\chi^0 = \chi^3$ and $\chi^\pm = (\chi^1 \mp i \chi^2)/\sqrt{2}$ for the neutral and charged wino states respectively.  Then $s_{00}$ ($s_{0\pm}$) provides the enhancement for a neutral wino initial state and a perturbative Feynman diagram involving neutral (charged) wino annihilation.  For a detailed discussion, see \emph{e.g.}~\cite{Beneke:2012tg, Cohen:2013ama}.  
 
For the LL line contribution, only the $s_{0\pm}$ contributes, since we are only matching the EFT to the full theory at tree level, which implies that the only non-zero diagrams are due to charged wino annihilation to $\gamma\,\gamma$ and $\gamma\,Z$.  However, in the endpoint region, the non-trivial combinations
\begin{align}
F_0 &= \frac43 \,\big|s_{00}\big|^2 + 2 \,\big|s_{0\pm}\big|^2 + {4\, \sqrt{2} \over 3}\, \Re\Big (s_{00} \,s^*_{0\pm} \Big) \,, \nonumber \\[3pt]
 F_1 &= - \frac43 \,\big|s_{00}\big|^2 + 2\,\big |s_{0\pm}\big|^2 - {4 \sqrt{2} \over 3}\, \Re\Big(s_{00}\, s^*_{0\pm}\Big ) \,,
\end{align}
also appear, where $\Re(\dots)$ gives the real part of the argument.  This occurs because three-body processes yielding $W^+\,W^-\,\gamma$ can now appear, and this channel is non-zero for both charged and neutral wino annihilation. Beyond LL the structure of both the line and endpoint spectrum contributions become more sophisticated, but the basic ingredients discussed here, and type of logarithms that are resummed, remain the same. For our analysis here we will make use of the full NLL results from \cite{us:NLL}.  These results for the case of wino DM can in principle be straightforwardly extended to other heavy WIMP DM.

The calculation presented here is based on an EFT expansion that requires that the resolution is much greater than $\mW/(2\,m_\DM)$. We find that our calculation is not reliable below $\sim 1$ TeV. For masses below this value we would have to match our prediction onto an EFT that is valid in the low mass region. This is beyond the scope of the current paper, and therefore we only consider $m_\DM\gtrsim1$ TeV. However, due to the high quality data in this low mass region, we believe it would be interesting to consider, and we intend to pursue this in future work. For recent EFT work relevant to DM masses below $1$ TeV, see~\cite{Beneke:2018ssm}.

%%%%%%%%%%%%%%%%%%%%%%%%%%%%%%%%%%%%%%%
\subsection{Interpreting the Signal Prediction}
\label{sec:interpretSignal}
%%%%%%%%%%%%%%%%%%%%%%%%%%%%%%%%%%%%%%%

Given this precise prediction for the gamma-ray spectrum resulting from wino annihilation, it is worth revisiting the procedure for converting this into a flux prediction that can be used to probe wino annihilation.  

In particular, one of our goals here is to understand the extent to which including the spectral shape -- the \emph{endpoint} spectrum in the results below -- impacts the limits one would set, when compared to the limits derived assuming that only the line contribution is relevant. To this end, we define the \emph{line} annihilation cross section $\sigma_\text{line}$ to be half the coefficient of $\delta\big(E - m_\DM\big)$ in the expression for $\text{d}\sigma/\text{d}E$. For example, in the LL case we take the differential spectrum in Eq.~(\ref{eq:resummed}) and derive that
$\sigma_\text{line} = 2 \,|s_{0\pm}|^2\, \hat \sigma_{\text{line}}$, where $\hat \sigma_\text{line}$ is defined in Eq.~(\ref{eq:partonicSigmaLine}), and we have used Eq.~(\ref{eq:z}) to convert $z$ into $E$.  We emphasize that the NLL result is used for all numerical analysis in what follows.  Our conventions are such that the contribution to $\text{d}\sigma/\text{d}E$ is normalized as  $2\, \sigma_\text{line}\, \delta\big(E - m_\DM\big)$, where the factor of 2 accounts for the presence of two photons in exclusive $\chi\, \chi \rightarrow \gamma\, \gamma$ annihilations.  For line events, we must also include the branching rate to $\gamma\,Z$ though, giving $\sigma_\text{line} = \sigma(\chi\,\chi \rightarrow \gamma\, \gamma) +(1/2)\,\sigma(\chi\,\chi\rightarrow\gamma\,Z)$. The analysis can then be interpreted as either a constraint on $\langle \sigma\,v \rangle_\text{line}$, or as a constraint on the DM profile using the predicted wino rate.

In order to include the endpoint spectrum in the analysis, we take the NLL analog of Eq.~(\ref{eq:resummed}), subtract the contribution proportional to the line $\delta\big(E - m_\DM\big)$, and normalize to $\sigma_\text{line}$.  This yields an analytic prediction for the endpoint spectral shape that we will refer to as $\big(\text{d}\mathcal{N}_{\gamma}(E)/\text{d}E\big)^\text{endpoint}$, specifically
\begin{align}
\bigg(\frac{\text{d}\sigma}{\text{d}E}\bigg)^{\text{NLL}} = \sigma_\text{line} \left[2\, \delta\big(E - m_\DM\big) + \left(\frac{\text{d}\mathcal{N}_\gamma}{\text{d}E}\right)^\text{endpoint}\right ]\,.\notag\\[3pt]
\end{align}
Note that the use of a new notation for the spectrum, $\text{d}\mathcal{N}_{\gamma}/\text{d}E$, rather than $\text{d} N_{\gamma}/\text{d}E$ as appeared in Eq.~\eqref{eqn:flux} is deliberate, and is designed to emphasize that we are using a spectrum normalized to the line cross section. 

\begin{table*}[tb]
\centering
\renewcommand{\arraystretch}{1.8}
\setlength{\tabcolsep}{4.5pt}
\setlength{\arrayrulewidth}{1.3pt}
\begin{tabular}{|r|c|c|c|c|c|c|c|}
\hline
\multirow{2}{*}{$i$-th ROI}\hspace{15pt} & \multirow{2}{*}{$\begin{array}{c}\text{Solid angle:}\\[-5pt] \Delta\Omega_i\,\,\, \big[10^{-4} \text{ sr}\big]\end{array}$} &  \multicolumn{6}{c|}{$J\text{-factor: } J_i \big(\Delta\Omega_i \big) \,\,\, \big[10^{20} \text{ GeV}^2\text{ cm}^{-5}\big]$}  \\[2pt]
\cline{3-8}
&  & Einasto & $r_\text{c} = 0.3  \text{ kpc}$ & $r_\text{c} = 0.5 \text{ kpc}$ & $r_\text{c} = 1 \text{ kpc}$ & $r_\text{c} = 3 \text{ kpc}$ &  $r_\text{c} = 5 \text{ kpc}$\\
\hline
1: $\big(\bar \theta_1= 0.3^\circ\big)$ & 0.31 & 3.76 & 1.08 & 0.60 & 0.23  & 0.035 &  0.012 \\
2: $\big(\bar \theta_2=0.4^\circ\big)$ & 0.50 & 5.16 & 1.14  & 0.97 & 0.38   &  0.056 & 0.019 \\
3: $\big(\bar \theta_3=0.5^\circ\big)$ & 0.69 & 6.15 & 2.40 & 1.34 & 0.52 & 0.078 &  0.026\\
4: $\big(\bar \theta_4=0.6^\circ\big)$ & 0.88 & 6.89 & 3.04 & 1.71 & 0.66  & 0.099  & 0.033\\
5: $\big(\bar \theta_5=0.7^\circ\big)$ & 1.08  &   7.45 & 3.67 & 2.07 & 0.81  & 0.12 & 0.040\\
6: $\big(\bar \theta_6=0.8^\circ\big)$ & 1.27 & 7.88 & 4.29 & 2.43 & 0.95  & 0.14 & 0.047\\
7: $\big(\bar \theta_7=0.9^\circ\big)$ & 1.46 & 8.20 & 4.90  & 2.79 & 1.09  & 0.16 &  0.055\\
%\hline
\vdots\hspace{31pt} & \vdots & \vdots &\vdots  & \vdots& \vdots  & \vdots &  \vdots\\
%\hline
37: $\big(\bar \theta_{37}=3.9^\circ\big)$ & 7.55  & 8.78  & 8.78   & 8.78  & 5.23   & 0.88 & 0.28  \\
\hline
\end{tabular}
\caption{ \small \label{tab:tableJ} Definitions of the $i$-th ROI together with the corresponding solid angle size, and value of the $J$-factor for the several DM profiles considered here.  For brevity, we only show the first $7$ ROIs, which are used in the $1^\circ$ analysis, and then skip to the 37th since this is the largest ROI considered in this work, and is used in the $4^\circ$ analysis.
}
\end{table*}

Finally, in addition to investigating the impact of the perturbative endpoint spectrum, we will also include the contribution to the gamma-ray flux from processes where the hard annihilation is to a Standard Model final state that generates a spectrum of photons due to its subsequent decay.  We refer to this as \emph{continuum} in the results below.  In the wino example, both the neutral and charged winos can annihilate to $W^\pm$ bosons, which then decay.  Note that both parton level processes must be included, since again they will be accessed by the mixing due to the Sommerfeld effect.  The Sommerfeld enhanced annihilation rate is then convolved with the final state photon spectrum provided by the PPPC 4 DM ID~\cite{Cirelli:2010xx}.  Following the same logic as with the endpoint spectrum, we take the PPPC results, and normalize them to $ \sigma_\text{line}$ in order to derive the continuum spectral shape, referred to as $\big(\text{d}\mathcal{N}_{\gamma}(E)/\text{d}E\big)^\text{continuum}$.

Now we are setup to compare the three levels of approximation -- $(i)$ line, $(ii)$  line + endpoint, and $(iii)$  line + endpoint + continuum.  Revisiting Eq.~(\ref{eqn:flux}), 
\begin{equation}
\hspace{-5pt}\frac{\text{d}\Phi^{\rm DM}_{\gamma}}{\text{d}E}\big(\Delta\Omega,E\big) =  \frac{\langle \sigma\,v \rangle_\text{line} \,J\big(\Delta\Omega\big)}{8\,\pi\,m_{\rm DM}^2} \left[\frac{\text{d}\mathcal{N}_\gamma(E)}{\text{d}E}\right] \,,\!\!
\label{eqn:fluxRevisited}
\end{equation}
where we have grouped the combination $\big(\langle \sigma\,v \rangle_\text{line} \,J\big(\Delta\Omega\big)\big)$ since this is the quantity that can be constrained using IACT data, and the bracketed term encodes the spectral shape,
\begin{align}
\notag\\[-5pt]
\left[\frac{\text{d}\mathcal{N}_{\gamma}(E)}{\text{d}E}\right] = \begin{cases} 
2\,\delta\big(E - m_\DM\big) & (i)
\\
\\
(i) + \left(\frac{\text{d}\mathcal{N}_{\gamma}(E)}{\text{d}E}\right)^\text{endpoint} & (ii)
\\
\\
(ii) + \left(\frac{\text{d}\mathcal{N}_{\gamma}(E)}{\text{d}E}\right)^\text{continuum}& (iii)
\\[5pt]
\end{cases}\,.
\end{align} 
Note that having a broad spectrum impacts the analysis, as it can lead to contamination outside of the energy window associated with a signal of a given mass.  Note that a benchmark choice $\langle \sigma\,v \rangle_\text{line} = 10^{-27} \text{cm}^3/\text{s}$ is used for many of the plots below, since this is about where the projected limit for a $3 \text{ TeV}$ wino will lie.  Now that we have a clear understanding of the various aspects of the signal prediction, we are ready to explain the procedure used for deriving our expected limits.

%%%%%%%%%%%%%%%%%%%%%%%%%%%%%%%%%%%%%%%
\section{Expected Sensitivity}
\label{sec:mockdata}
%%%%%%%%%%%%%%%%%%%%%%%%%%%%%%%%%%%%%%%

Now that we have reviewed the experimental setting, and the theoretical prediction, we have everything we need to explain the data analysis strategy.  First, we will introduce many regions of interest so that we can explore optimizing the search strategy.  We show how to calculate the number of expected events for both signal and background, followed by an explanation of the 2D-binned likelihood procedure used to calculate the expected sensitivity to wino DM for a H.E.S.S.-like data set.

%%%%%%%%%%%%%%%%%%%%%%%%%%%%%%%%%%%%%%%
\subsection{Definition of Regions of Interest}
%%%%%%%%%%%%%%%%%%%%%%%%%%%%%%%%%%%%%%%

\begin{figure*}[t] 
\begin{center}
\includegraphics[trim=0cm -2.5cm 0cm 0cm,clip=true,width=0.5\textwidth]{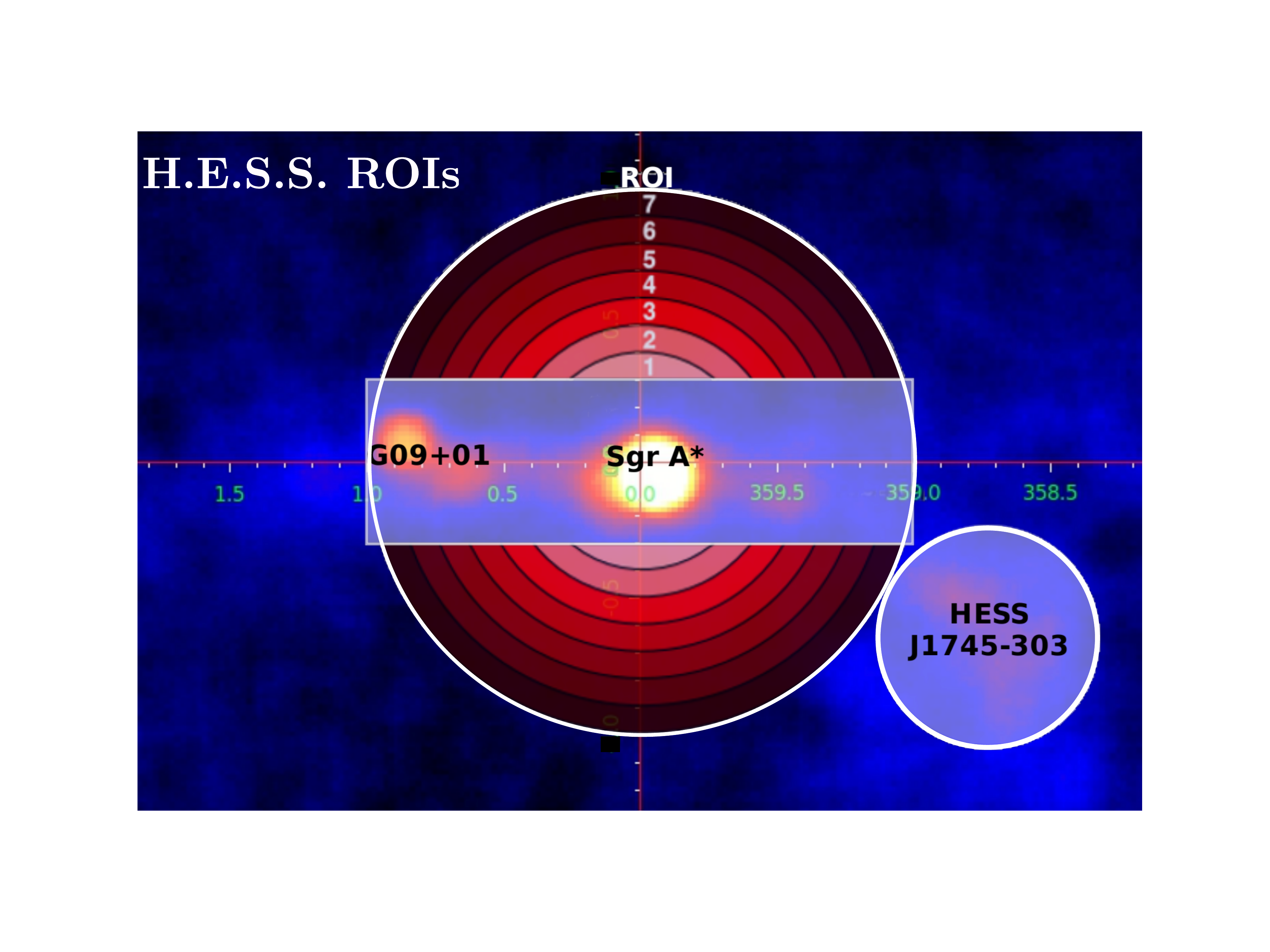}
\hfill
\includegraphics[width=0.48\textwidth]{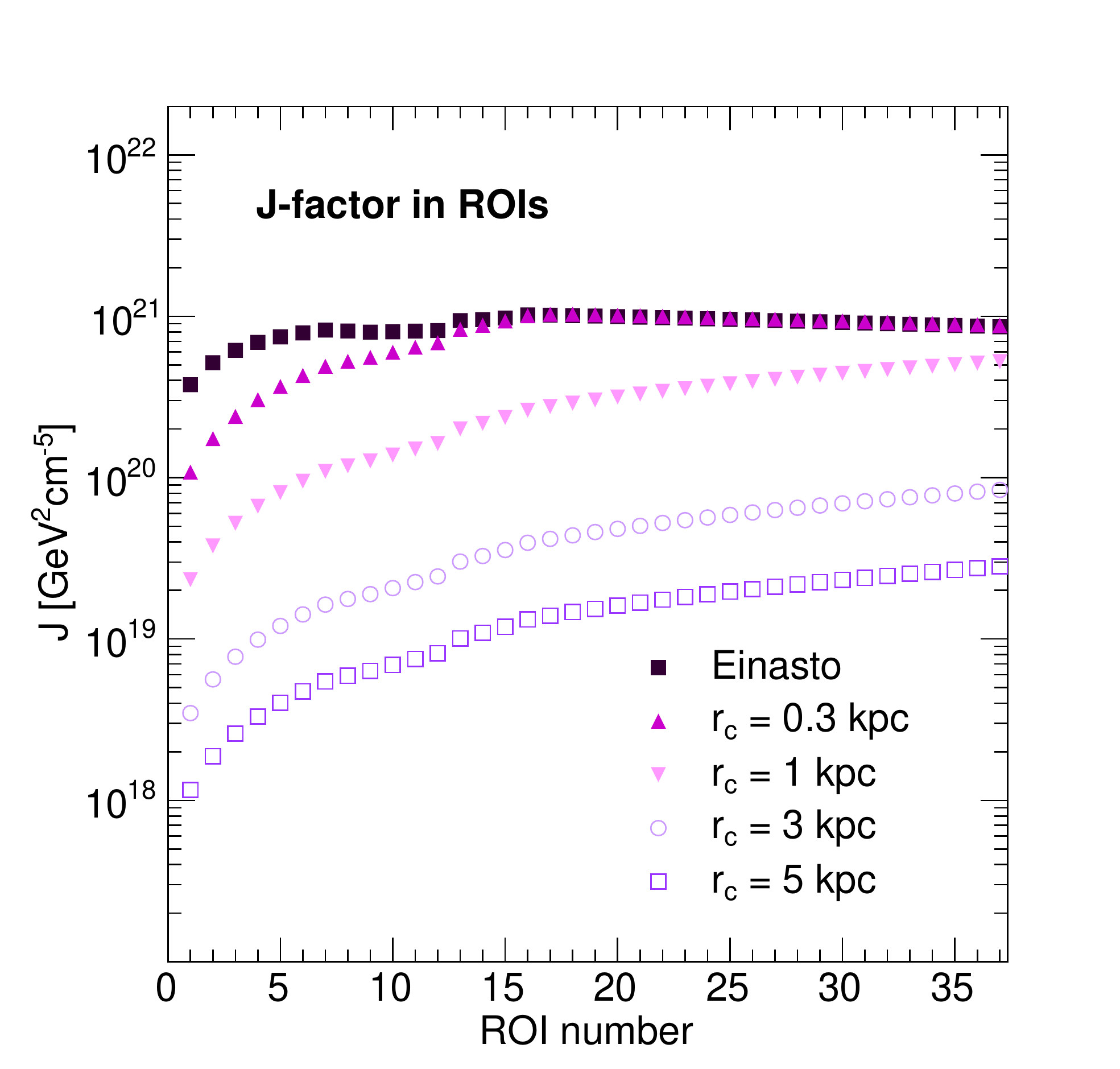}
\caption{{\it Left:} An illustration of the seven innermost ROIs. Figure adapted from~\cite{Lefranc:2015vza}. The three brightest gamma ray sources are labelled.  We mask the boxed region around the Galactic plane. When the ROI region becomes large enough for this to become relevant, we also exclude a disk centered on the position of the source HESS J1745-303 as discussed in the text.
{\it Right:} $J$-factors integrated over the ROI solid angle as a function of the ROI number, for several DM profiles studied here.}
\label{fig:ROIandJFactor}
\end{center}
\end{figure*} 

The area of the sky that will be targeted to extract the signal follows from~\cite{Abdallah:2016ygi,Abdallah:2018qtu}.  This is the ROI, and we will often use a shorthand and refer to this as the ON region.  It is defined as a circle with a one-degree radius centered on the GC. In order to best exploit the spatial behavior of a DM signal as compared to background, this region is then split in several sub-ROIs.

As mentioned in Sec.~\ref{sec:GCTeV}, the GC is a very crowded environment that produces VHE gamma-rays through a variety of mechanisms. In order to avoid the need to model the contamination from known astrophysical sources of gamma-rays, the corresponding regions of the sky are excluded from the ON regions. In particular, a box with longitudes $|\ell|<$1.5$^{\circ}$ and latitudes $|b|<$0.3$^{\circ}$ in Galactic coordinates, and a disk of 0.4$^{\circ}$ radius centered at $(l,b)=(-1.29^{\circ},-0.64^{\circ})$ are masked in order to exclude the diffuse emission along the Galactic plane and the known source HESS J1745-303.

The $i$-th ON regions are concentric rings of $i$-th aperture $\bar \theta_i$ ranging from  0.3$^{\circ}$ to 0.9$^{\circ}$ and constant width $\Delta\theta =0.1^{\circ}$.
The solid angle of the $i$-th ROI is defined as
\begin{equation}\label{eq:DeltaOmega}
\Delta\Omega_{i} = \Delta\Omega_i^{\rm ring} -   \Delta\Omega_i^{\rm excluded} \,,
\end{equation}
where
\begin{equation}
\displaystyle \Delta\Omega^{\rm ring}_i= 2\,\pi\,\int_{\bar\theta_i}^{\bar\theta_i+\Delta\theta} \text{d} \theta\,  \sin\theta \,,
\end{equation}
is the solid angle of the $i$-th ring and
\begin{equation}
\displaystyle \Delta\Omega^{\text{excluded}}_i= 4\int_{\bar\theta_i}^{\bar\theta_i+\Delta\theta}  \text{d} \ell  \int_{0}^{b_{\text{max}}} \text{d} b \, \cos b \,,
\end{equation}
is the solid angle of the $i$-th excluded region. The corresponding $J$-factor in the $i$-th ROI is then defined as $J\big(\Delta\Omega_i\big) = J\big(\Delta\Omega_i^{\rm ring}\big) - J\big(\Delta\Omega_i^{\rm excluded}\big)$.  Additionally, the solid angle of any ring that intersects the HESS J1745-303 exclusion region is also removed.

The left panel of Fig.~\ref{fig:ROIandJFactor} illustrates the first seven ROIs, and Table~\ref{tab:tableJ} gives explicit values of the solid angle size  $\Delta\Omega_i$ and $J$-factor assuming several of the DM profiles considered in this study, for the first 7 and 37th ROIs.  The $i$-th $J$-factor integrated over the solid angle size $\Delta\Omega_i$ is plotted in the right panel of Fig.~\ref{fig:ROIandJFactor} for all 37 ROIs.  As expected, if a given ROI lies within the core radius, the $J$-factors simply scales with the solid angle size $\Delta\Omega$.  The increase of $J \big(\Delta\Omega \big)$ at the 12th ROI corresponds to an increase in the integration region for the $J$-factor computation where the Galactic plane is no longer excluded. Beyond the 12th ROI the $J$-factor is given by $J\big(\Delta\Omega_{\rm ring}\big)$.

\begin{figure*}[t!] 
\begin{center}
\includegraphics[width=0.48\textwidth]{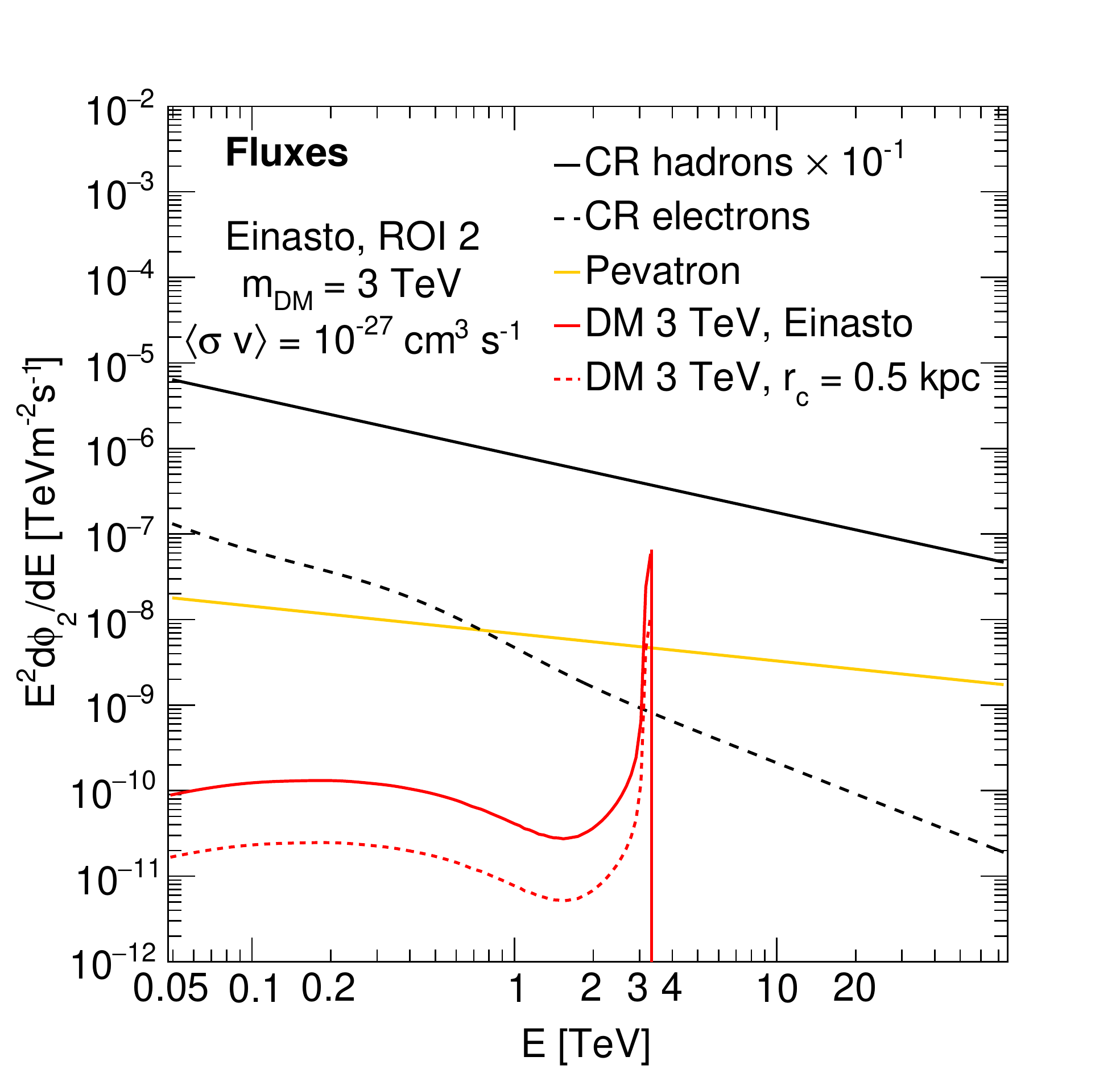}
\hfill
\includegraphics[width=0.48\textwidth]{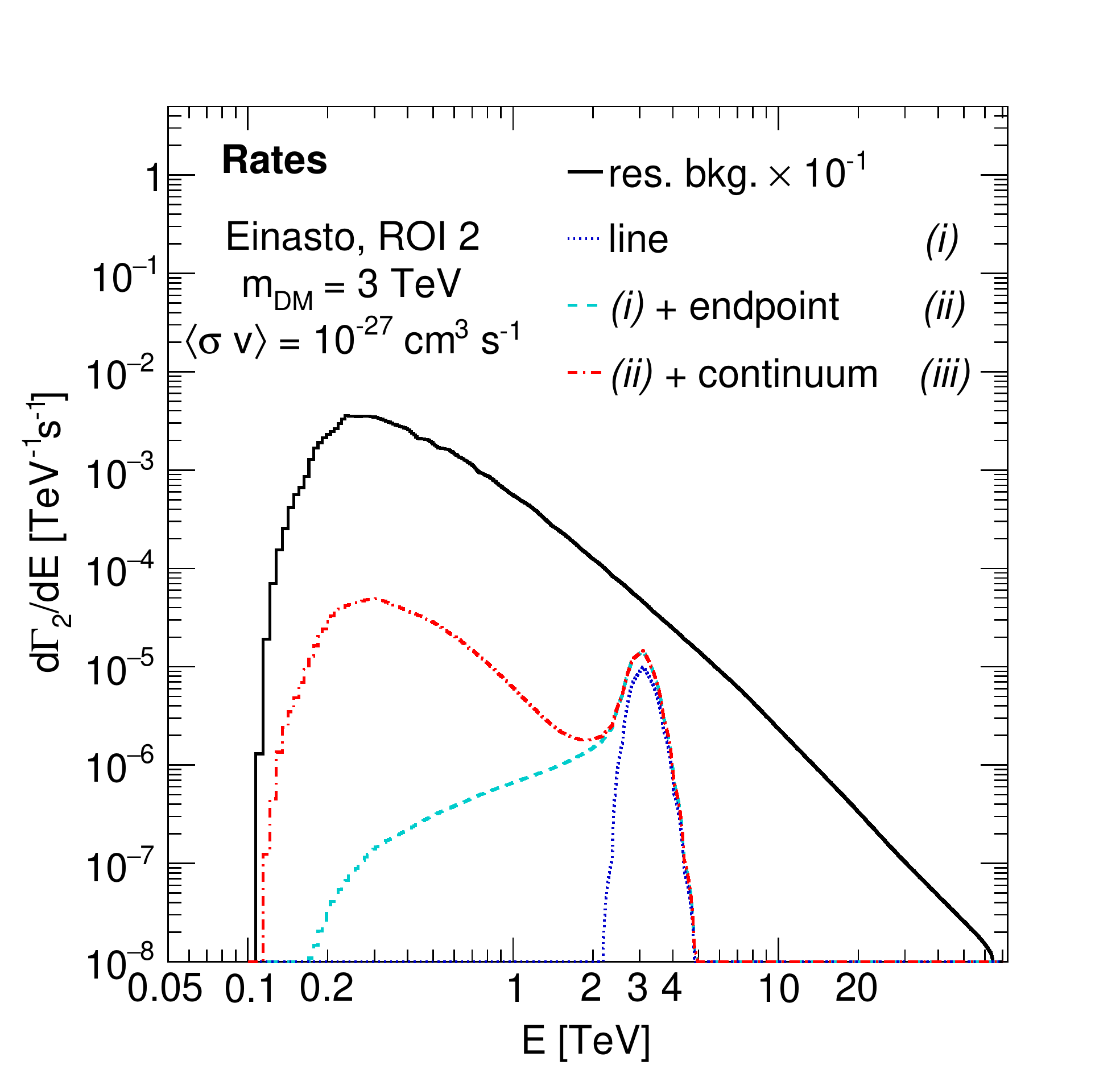}
\caption{{\it Left:}  Expected differential flux in ROI 2 for a 3 TeV wino annihilating with an annihilation cross section $\langle \sigma\,v \rangle_\text{line} = 10^{-27} \text{ cm}^3 \text{ s}^{-1}$, for the Einasto (solid red line) and 0.5 kpc cored (dashed red line) DM profiles. Only the continuum and endpoint components are displayed. The CR hadron  (proton + nuclei) flux (solid black line) is plotted together with the CR electron flux (black dotted line) and the diffuse flux from the H.E.S.S. Pevatron (orange solid line). {\it Right:} Expected differential count rate as a function of energy for signal and background in ROI 2 for a wino mass of 3~TeV with an annihilation cross section $\langle \sigma\,v \rangle_\text{line} = 10^{-27}\text{ cm}^3 \text{ s}^{-1}$. For the signal, the differential count rates are given for the line (dotted blue line), the line + endpoint (dashed cyan line), and the line + endpoint + continuum (dashed-dotted red line). The residual background is plotted as a solid black line. Here the effect of the convolution with the energy resolution is included, which is not the case in the left figure.}
\label{fig:signal}
\end{center}
\end{figure*}

The background in the ON region is determined from regions in the sky, hereafter referred to as OFF regions, using the reflected background method~\cite{Abdallah:2018qtu}.  For each observation and each ON region, the corresponding OFF region is defined as the area of the sky symmetric to the ON region with respect to the pointing position. In this way, the ON and OFF regions have the same exposure, acceptance, solid angle size and observational conditions, \emph{i.e.}, no further correction is needed to compare the ON and OFF regions.
The procedure is then repeated for all the ROIs and all the observations at different pointing positions. Such a method has proven to be efficient in the case of cuspy DM profiles for which a DM gradient is expected between 
the ON and OFF regions~\cite{Abdallah:2016ygi,Abdallah:2018qtu}, or DM profiles where the core extends below about 100 pc. 
For profiles with kpc-sized cores, a different approach is required. One can model the expected background using dedicated simulations of the observational and instrumental conditions during data taking~\cite{Holler:2017ynz}. 
Alternatively, the background can be measured utilizing dedicated observations taken in the closest possible observational conditions as the signal measurement observations. In particular, the zenith angle and offset between the ROI and pointing position should be chosen to be similar to the observation conditions of the ON region. However, this strategy implies that the overall observation time is doubled, and additional corrections may be required to account for changes in the observational and instrumental conditions. The systematic uncertainties could be larger than in the case of precise Monte Carlo simulations of the residual background for each observation run~\cite{Holler:2017ynz}.   Therefore, we use the strategy of simulating the expected background in the observation conditions in what follows, leaving a detailed exploration of the data driven approach for future work.

%%%%%%%%%%%%%%%%%%%%%%%%%%%%%%%%%%%%%%%
\subsection{Number of Signal and Background Events}
%%%%%%%%%%%%%%%%%%%%%%%%%%%%%%%%%%%%%%%

\begin{figure*}[tbh] 
\begin{center}
\includegraphics[width=0.48\textwidth]{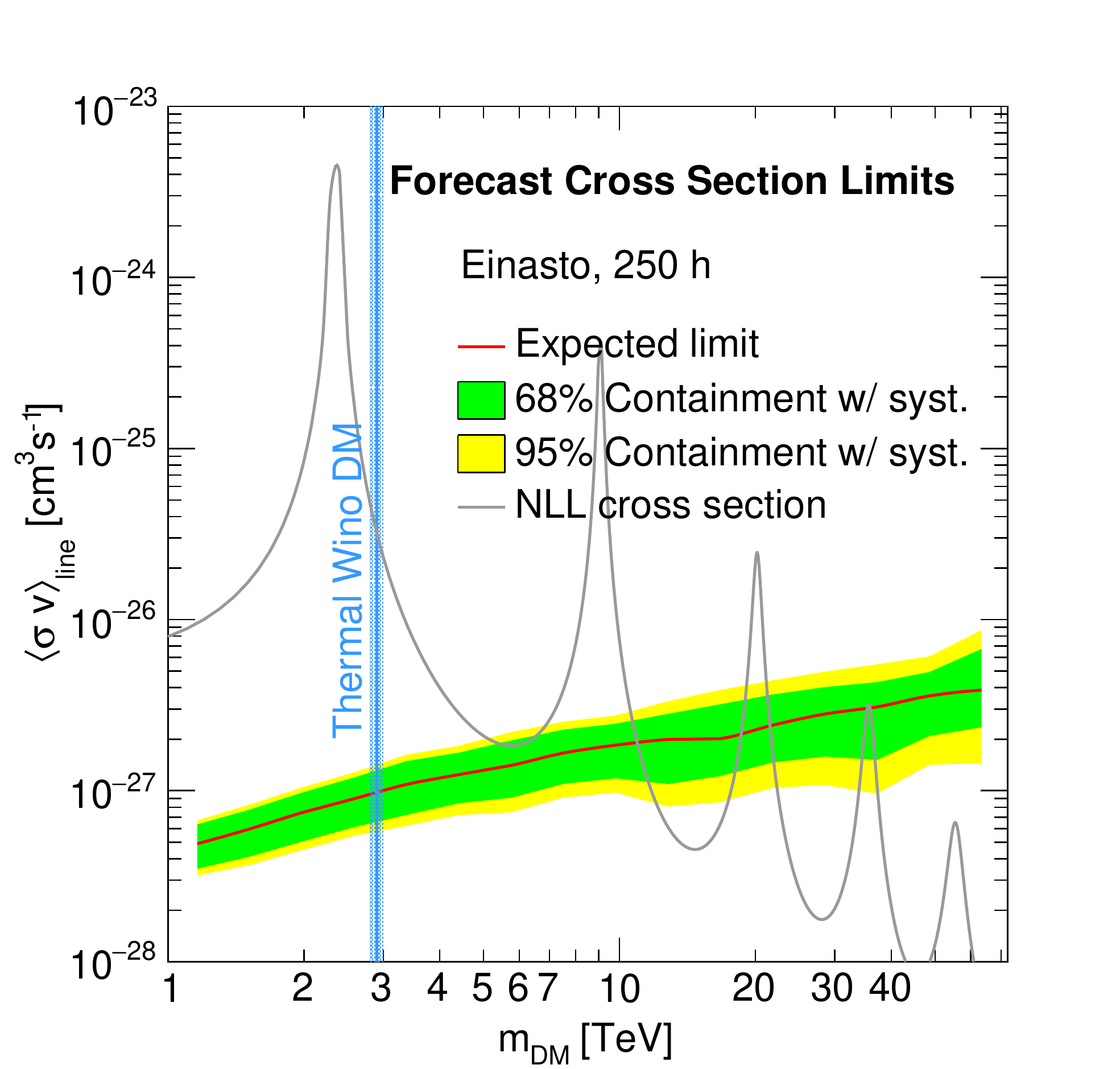}
\hfill
\includegraphics[width=0.48\textwidth]{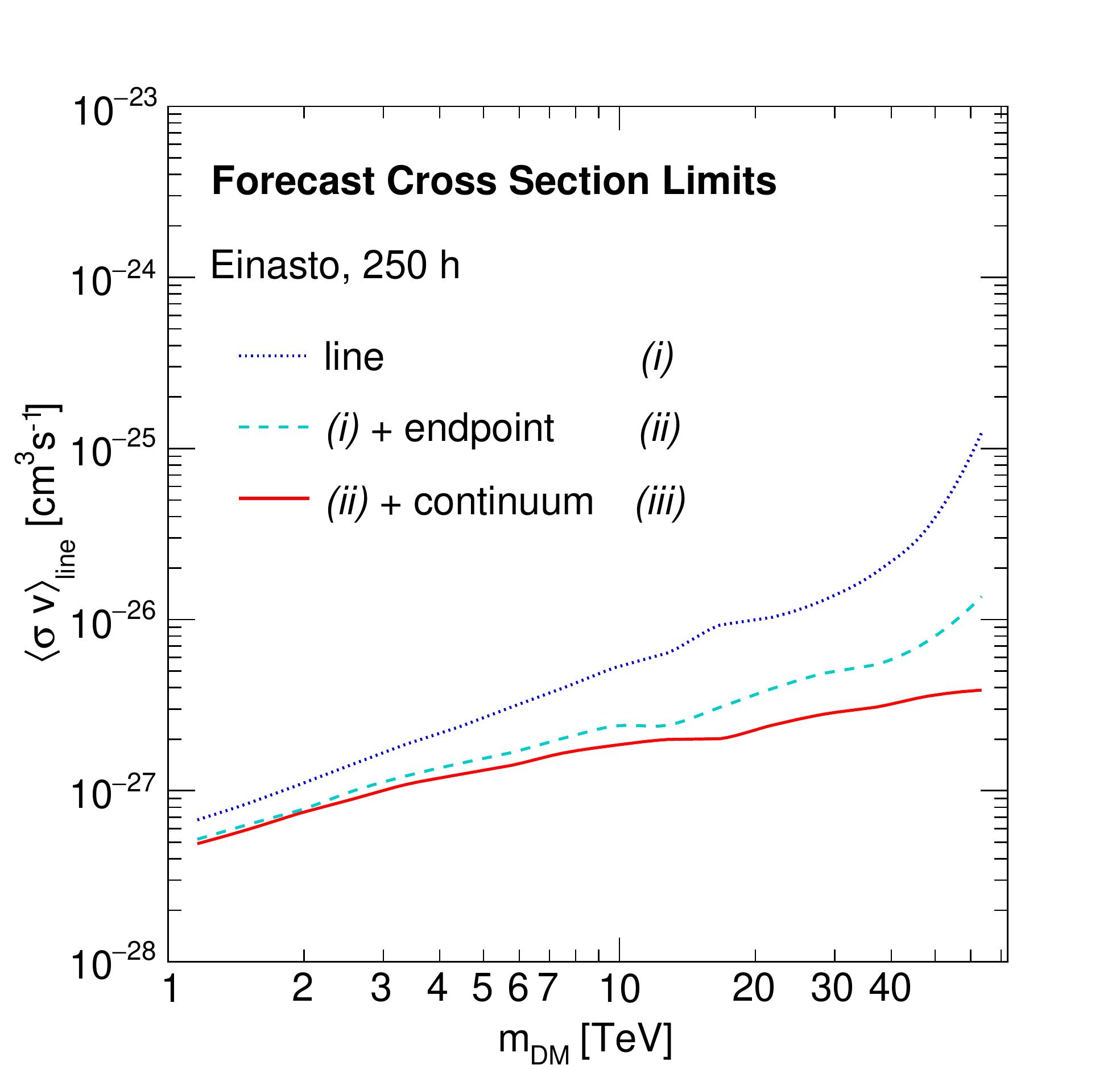}
\caption{95\% C.L. mean expected upper limits on the thermally-averaged velocity-weighted annihilation cross section $\langle \sigma\,v \rangle_\text{line}$ as a function of the DM mass $m_\DM$.  
{\it Left:} The mean expected limits together with statistical 68 and 95\% containment bands including the systematic and theoretical uncertainties. The mass corresponding to a thermally-produced wino DM $m_\DM = 2.9 \pm 0.1\text{ TeV}$~\cite{Beneke:2016ync} is shown as a light blue vertical band. The NLL cross section for wino DM is shown in gray. {\it Right:} The limits are given assuming the line-only (blue dotted line), a line + endpoint contribution (cyan dashed line), and line + endpoint + continuum spectrum (red solid line).}
\label{fig:compare}
\end{center}
\end{figure*}

Assuming a DM annihilation channel and a DM density profile, the expected number of signal gamma-rays $N_{\text{S},ij}$  in the $i$-th ROI and $j$-th energy bin can be written as 
\begin{equation}
N_{\text{S},ij} = T_{\text{obs},i} \int_{E_j-\Delta E_j/2}^{E_j+\Delta E_j/2}\text{d}E'\,\frac{\text{d}\Gamma_{\text{S},ij}}{\text{d}E'} \,,
\end{equation}
where $T_{\text{obs},i}$ is the observation time in seconds, and
\begin{equation}
 \frac{\text{d}\Gamma_{\text{S},ij}}{\text{d}E} = \int_{-\infty}^{\infty}\!\text{d}E'\,\frac{\text{d}\Phi^{\DM}_{\gamma,ij}}{\text{d}E'}\big(\Delta\Omega,E'\big) \,A_{\text{eff}}^{\gamma}\big(E'\big)\, G\big(E_j-E'\big) \, ,\\[5pt]
\label{eq:NDM}
\end{equation}
where $\text{d}\Phi^{\DM}_{\gamma}/{\text{d}E'}$ is the energy-differential self-annihilation spectrum defined in 
Eq.~(\ref{eqn:flux}), see Sec.~\ref{sec:interpretSignal} for a discussion of the conventions used when comparing the different theory approximations.
  The instrument response function is encoded in the following two terms: $A_{\rm eff}^{\gamma}$ is the energy-dependent acceptance to gamma-rays; and $G$ is a Gaussian that models the finite energy resolution of the instrument.  The gamma-ray acceptance for H.E.S.S.-I observations of the GC region is extracted from~\cite{lefra2016}, and following~\cite{Abdallah:2018qtu}, a width of $\sigma/E$ of 10\% is used in $G$.

The backgrounds for gamma-ray measurements are dominated by CR protons and nuclei reaching the Earth's atmosphere. While the majority of these CRs can be efficiently rejected using both a shower shape parameter measurement and a stereoscopic view of the events seen by the IACT~\cite{Aharonian:2006pe}, a fraction of them cannot be distinguished from showers initiated by VHE gamma-rays, implying that there is a limit to our ability to reject the CR background. Following~\cite{2013APh43171B}, 
the overall CR flux includes the flux measurements of CR protons and helium as well as electrons and positrons. Given the finite discrimination between gamma-rays and CR protons and nuclei, a constant rejection factor of 10 is assumed. On the other hand, showers initiated by electrons and positrons may be distinguished from those from gamma-rays using the reconstructed primary interaction depth on the incident particle in the atmosphere -- here we use a constant rejection factor of 1.

Following Eq.~(\ref{eq:NDM}), the expected number of background photons $N_{\rm B}$ from the residual CR background can be computed using 
\begin{equation}
N_{\text{B},ij} = T_{\text{obs},i} \int_{E_j-\Delta E_j/2}^{E_j+\Delta E_j/2} \text{d}E' \,\frac{\text{d}\Gamma_{\text{B},ij}}{\text{d}E'} \,,
\end{equation}
where
\begin{equation}
 \frac{\text{d}\Gamma_{\text{B},ij}}{\text{d}E} = \int_{-\infty}^{\infty}\!\text{d}E'\,\frac{\text{d}\Phi^{\text{CR}}_{\gamma,ij}}{\text{d}E'}\big(\Delta\Omega,E'\big) A_{\text{eff}}^{\text{CR}}\big(E'\big)\, G\big(E_j-E'\big) \,, \\[5pt]
\label{eq:NBck}
\end{equation}
where  $\text{d}\Phi^{\text{CR}}_{\gamma}/{\text{d}E'}$ is the overall CR flux, and $A_{\rm eff}^{\rm CR}$ is the energy-dependent acceptance for CRs. Here we assume that $A_{\rm eff}^{\rm CR} = \epsilon_{\rm CR}\, A_{\rm eff}^{\gamma}$, where  $\epsilon_{\rm CR}$ is the CR efficiency and is taken to be 10\% over the full energy range considered here. 
Additionally, a known gamma-ray contamination due to the Pevatron in the GC~\cite{Abramowski:2016mir} is added to the first two ROIs. 
Refining our mock description of the H.E.S.S. response function in the GC region would require a full simulation of the instrument, which is beyond the scope of the present study.

\begin{figure}[tbh] 
\begin{center}
\includegraphics[width=0.48\textwidth]{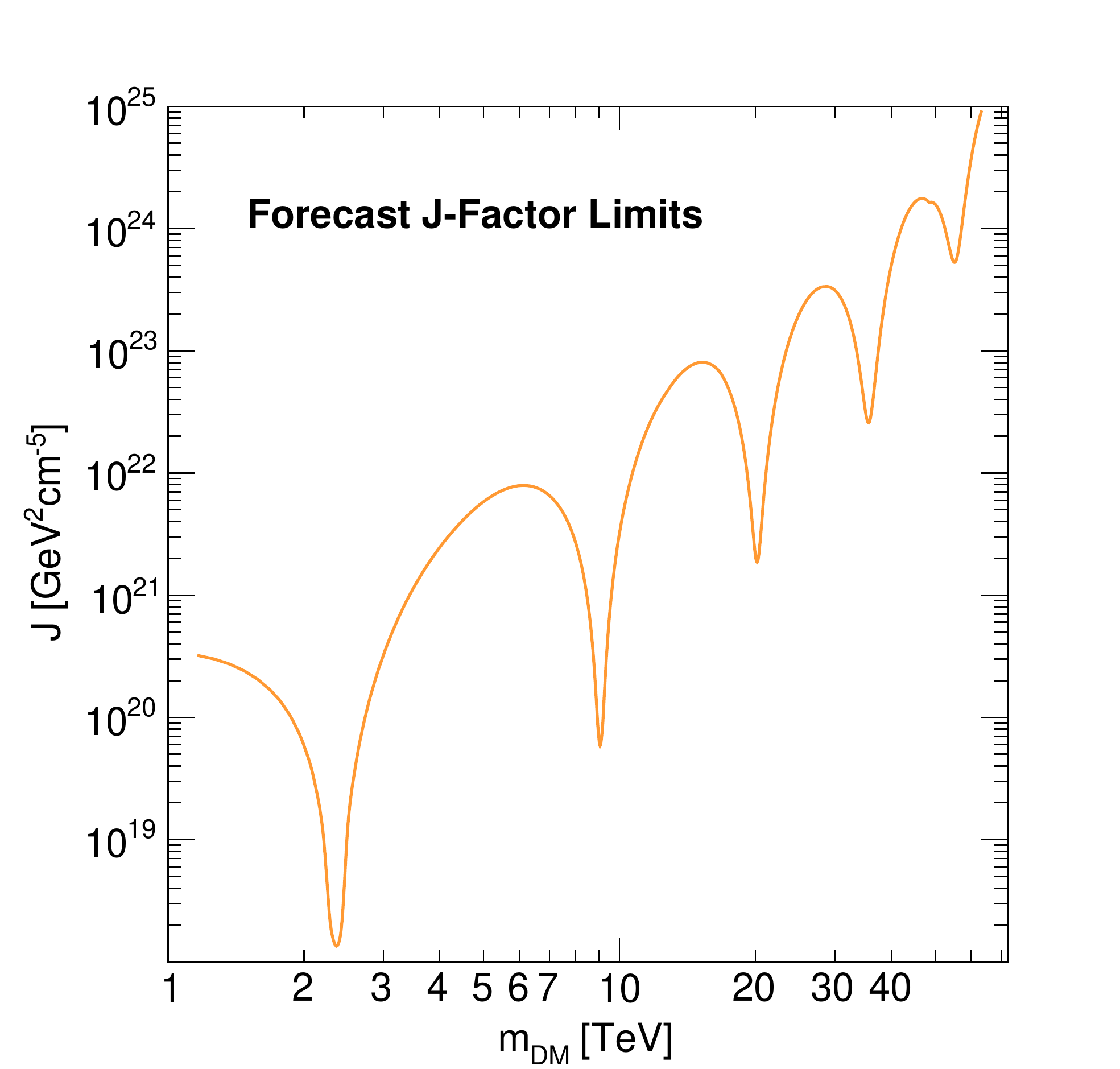}
\caption{Mean expected upper limits at 95\% C.L. on the $J$-factor as a function of the DM mass $m_\DM$. }
\label{fig:Jcorelimit}
\end{center}
\end{figure} 

The left panel of Fig.~\ref{fig:signal} shows the expected fluxes in ROI 2 for a 3 TeV wino with a velocity-weighted annihilation cross section of $\langle \sigma\,v \rangle_\text{line} = 10^{-27} \text{ cm}^3\text{ s}^{-1}$ for both the Einasto and 0.5 kpc-size cored DM profiles. The CR proton plus nucleus flux and the CR electron flux are also plotted.  The gamma-ray flux from the Pevatron detected by H.E.S.S. is shown for the same ROI.
The right panel of Fig.~\ref{fig:signal} shows the differential count rate of signal  in ROI 2 for a 3 TeV wino with $\langle \sigma\,v \rangle_\text{line}$ = 10$^{-27}$ cm$^3$s$^{-1}$, and the residual background, respectively.   This provides some intuition for how the signal and background rates compare in the GC.  The task of our statistical framework is then to distinguish these two sources of gamma-rays from each other such that annihilating winos could be discovered using H.E.S.S..

%%%%%%%%%%%%%%%%%%%%%%%%%%%%%%%%%%%%%%%
\subsection{Statistical Procedure}
%%%%%%%%%%%%%%%%%%%%%%%%%%%%%%%%%%%%%%%

Our statistical procedure utilizes the maximal likelihood approach.  The 2D-binned likelihood function in the $i$-th spatial bin (corresponding to the annular ROI) and in the $j$-th energy bin is a product of the Poisson probabilities in the ON and OFF regions. Denoting $\text{Pois}(\lambda,k) = e^{-\lambda}\,\lambda^k/k!$, we have
\begin{align}
&\mathcal{L}_{ij}\big(N^{\rm ON},N^{\rm OFF},\alpha \big|N_{\rm S}^{\text{ON}},N_{\rm S}^\text{OFF},N_{\rm B}\big)= \\[5pt]
&\text{Pois}(N_{\text{S},ij}^\text{ON}+N_{\text{B},ij}, N^{\text{ON}}_{ij} )     \text{Pois}( N^\text{OFF}_{\text{S},ij}+\alpha_i\, N_{\text{B},ij}, N^{\text{OFF}}_{ij} )\,,\nonumber
\end{align}
where again the subscripts denote the $i$-th ROI and the $j$-th energy bin.  Here $N^{\rm ON}$ and $N^{\rm OFF}$ are the measured count numbers in the ON and OFF regions respectively, while $N_{\rm S}^{\rm ON}$ and $N_{\rm S}^{\rm OFF}$ correspond to the expected DM signal in the ON and OFF regions, respectively. $N_{\rm B}$ is the count number of the expected residual background, which is the same in both the ON and OFF region. The $\alpha_i$ parameter defined as  $\alpha_i \equiv \Delta \Omega^{\rm OFF}/\Delta \Omega^{\rm ON}$ corresponds to the ratio between the solid angle of the 
OFF region to that in corresponding ON region. Here we take $\alpha_i= 1$ and $N_{\rm S}^\text{OFF}$ = 0, since the OFF count numbers are obtained from a simulation of the residual background. The full likelihood is then given by the product of the likelihoods for each bin in the given ROI
\begin{align}
\mathcal{L} = \prod\limits_{i \in \text{ROI}} \prod\limits_j \mathcal{L}_{ij}\,.
\end{align}

Through this likelihood, we can use the data to compute a limit on the following combination of parameters as a function of mass,
\begin{align}
\kappa = \langle \sigma\,v \rangle_\text{line} \times J(\Delta\Omega)\,.
\end{align}
In detail, we do this using the likelihood ratio test statistic defined by
\begin{align}
{\rm TS}(m_\DM)=-2 \ln\left[\frac{\mathcal{L}\big(m_\DM,\kappa \big)}{\mathcal{L}\big(m_\DM, \widehat{\kappa} \big)}\right]\,,
\end{align} 
where $\widehat{\kappa}$ denotes the value of $\kappa$ which maximizes the likelihood for the given DM mass. Then, for a given model prediction for $J(\Delta\Omega)$, we can convert this constraint into one on $\langle \sigma\,v \rangle_\text{line}$, or similarly a constraint on the $J$-factor for a given cross-section model.
The TS distribution follows an approximate $\chi^2$ distribution with one degree of freedom. Values of TS equal to 2.71 provides one-sided upper limits on $\langle \sigma\,v \rangle_\text{line}$ at a 95\% Confidence Level (C.L.). 100 Poisson realizations of the expected signal and of the expected background are performed. For each realization, the likelihood ratio test statistic is computed to obtain the expected limit.  We show the mean expected limit on $\langle \sigma\,v \rangle_\text{line}$, together with bands calculated as follows.  We derive the standard deviation of the distribution of the $\langle \sigma\,v \rangle_\text{line}$ values obtained from our 100 realizations and combine this linearly with the systematic and theoretical uncertainties. This value is used to determine the one and two sigma deviations from the mean, thereby providing the bands shown below in Fig.~\ref{fig:compare}.

%%%%%%%%%%%%%%%%%%%%%%%%%%%%%%%%%%%%%%%
\section{Results and Prospects}
\label{sec:results}
%%%%%%%%%%%%%%%%%%%%%%%%%%%%%%%%%%%%%%%

\begin{figure*}[tbh] 
\begin{center}
\includegraphics[width=0.48\textwidth]{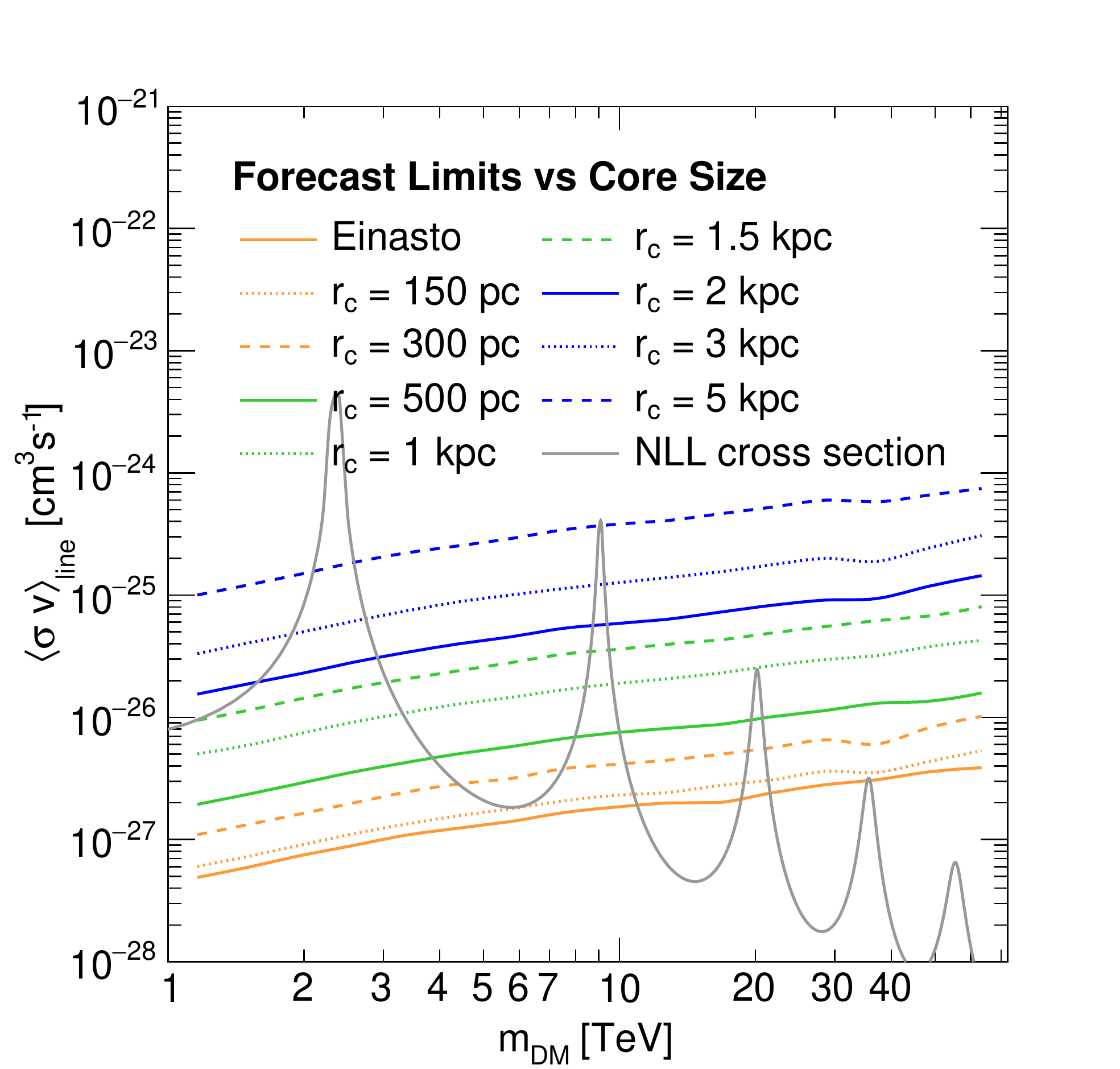}
\hfill
\includegraphics[width=0.48\textwidth]{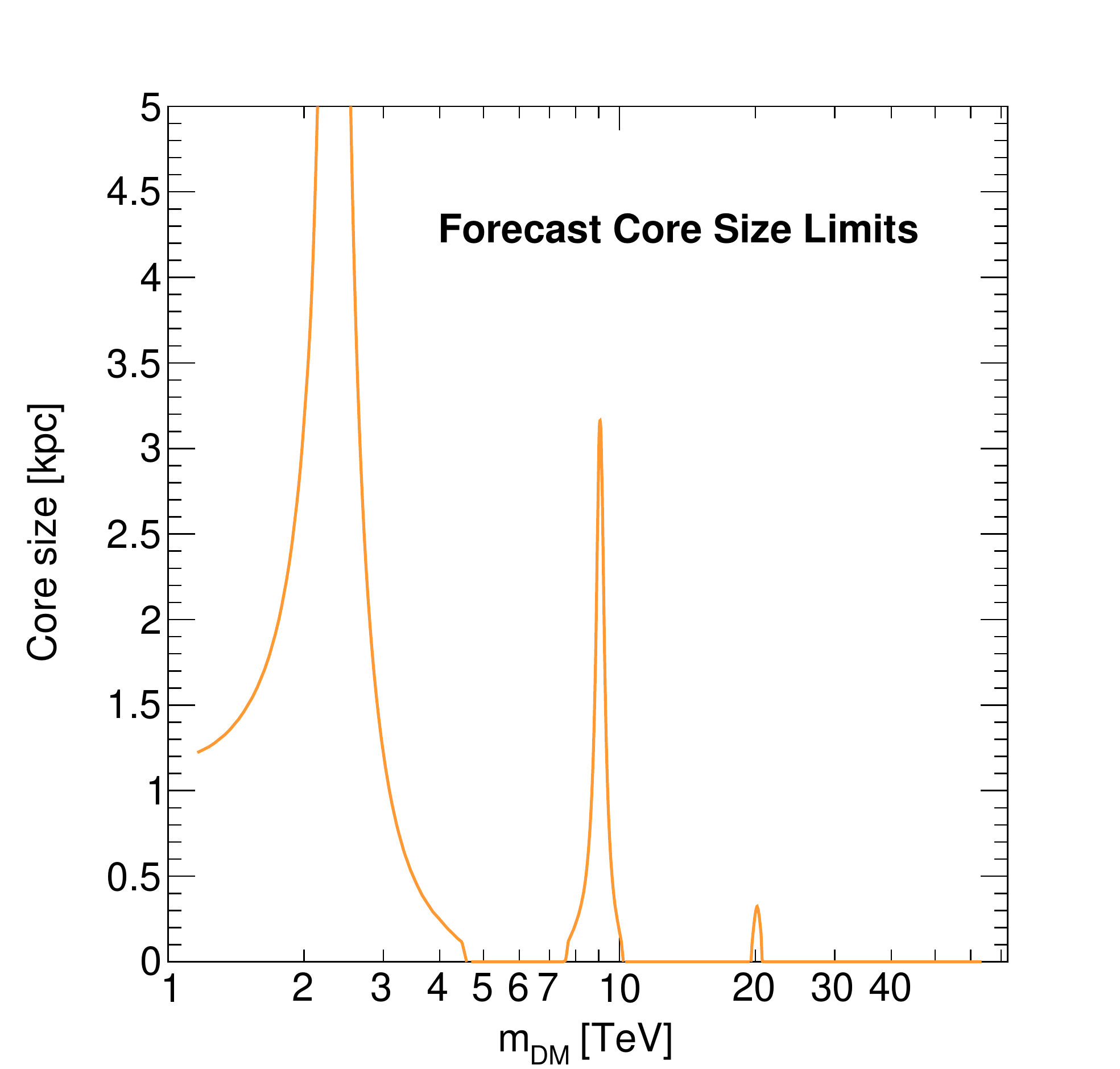}
\caption{ {\it Left:} Mean expected upper limits at 95\% C.L.  on the wino annihilation cross section $\langle \sigma\,v \rangle_\text{line}$ as a function of the DM mass $m_\DM$ assuming the full line + endpoint + continuum spectrum considering ROIs up to $1^\circ$, where the profile is varied as labeled in the legend. The NLL cross section for wino DM is shown in gray.
{\it Right:} Mean expected lower limits at 95\% C.L. on the DM core size as a function of the DM mass $m_\DM$, required to save the wino model.}
\label{fig:corelimit}
\end{center}
\end{figure*}

Fig.~\ref{fig:compare} shows the 95\% C.L. upper limits on the velocity-weighted annihilation cross section $\langle \sigma\,v \rangle_\text{line}$ as a function of the wino mass assuming ROIs up to 1$^{\circ}$. The left panel of Fig.~\ref{fig:compare} shows the 95\% C.L. mean expected upper limits together with the 68\% and 95\% containment bands, for the Einasto DM profile. The containment bands include the systematic uncertainties for DM searches in the GC region as estimated by  H.E.S.S. in~\cite{Abdallah:2018qtu}.  The main sources of error come from the small dependence of energy resolution on the observation conditions, the imperfect knowledge of the energy scale, and the influence of the variation of the night sky background (NSB)\footnote{The NSB corresponds to the optical photons emitted from bright stars in the field of view of the telescope.} in the field of view on the event count measurements. 
The inhomogeneous NSB rate in the field of view of the GC observations implies a a shift in the limits from a few percents up to 60\% depending on the DM mass range. The systematic uncertainties can likely be lowered down using �accurate simulations of the residual background~\cite{Holler:2017ynz} in the same instrumental and observational conditions of the ROIs in the GC.  In particular, the measured NSB rate in each pixel of the region of interest can be used allowing for further subtraction of this component. As pointed out in~\cite{Abdallah:2018qtu}, 
a systematic uncertainty in the energy scale of 10\% shifts the limits by up to 15\%. The weak dependency of the energy resolution on the observational condition is a subdominant source of systematic uncertainty. An artificial deterioration of the energy resolution by a factor two would only weaken the limits by 25\%.
 The theoretical uncertainty from the full wino spectrum is taken into account, it yields an uncertainty on the limits from a few percents up to $\sim10\%$ at the highest masses.  The NLL theory cross section for a wino as a function of $m_\DM$ is shown in gray. Further sophisticated DM searches in the GC will tackle the experimental systematic uncertainties in this complex environment by a careful consideration of the instrumental and observational conditions in the analyses. Provided that the sources of systematic uncertainties can be controlled up to a level of the theoretical uncertainty a higher precision in the theoretical computation may be relevant.

\begin{figure*}[tbh] 
\begin{center}
\includegraphics[width=0.48\textwidth]{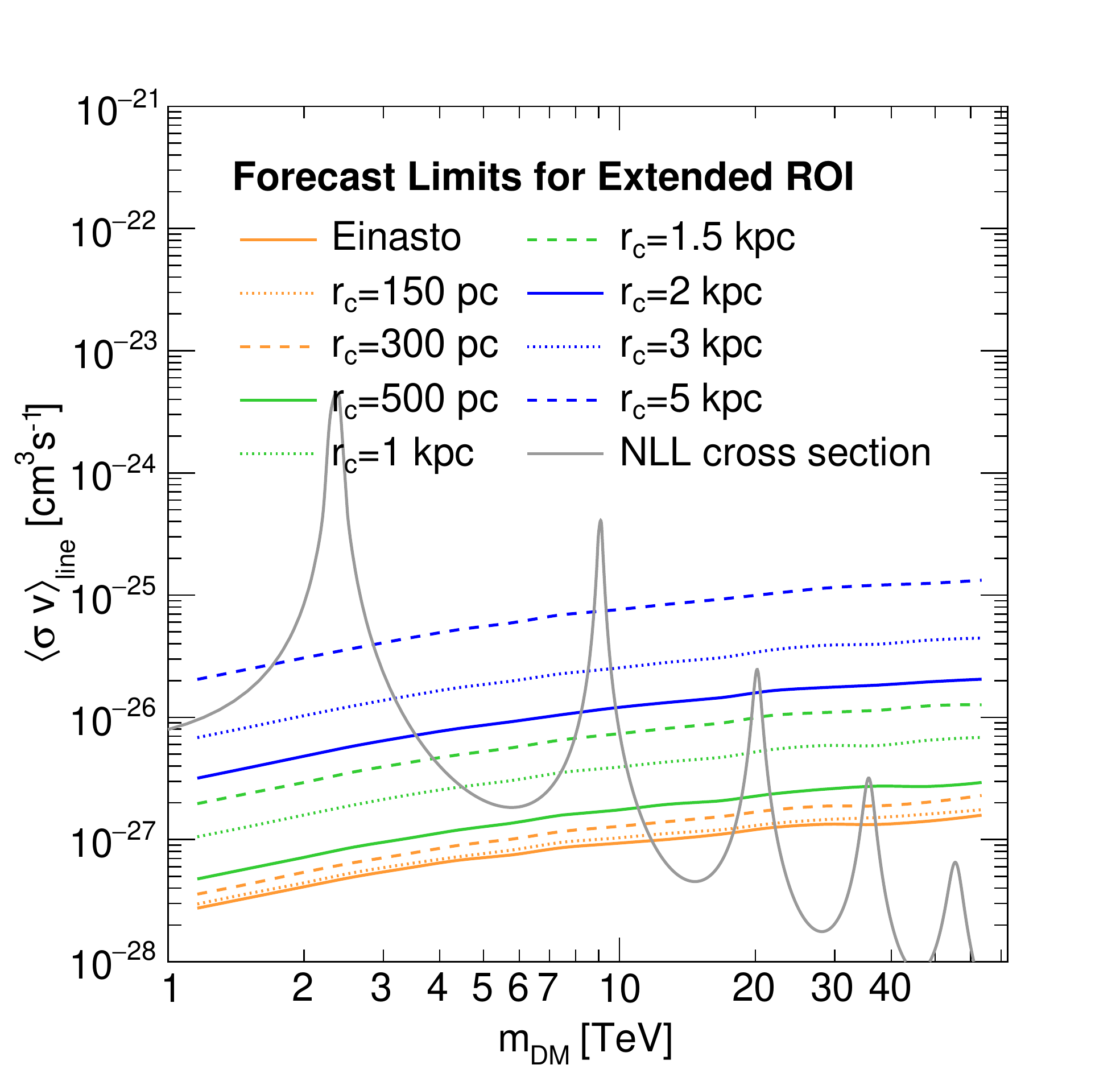}
\hfill
\includegraphics[width=0.48\textwidth]{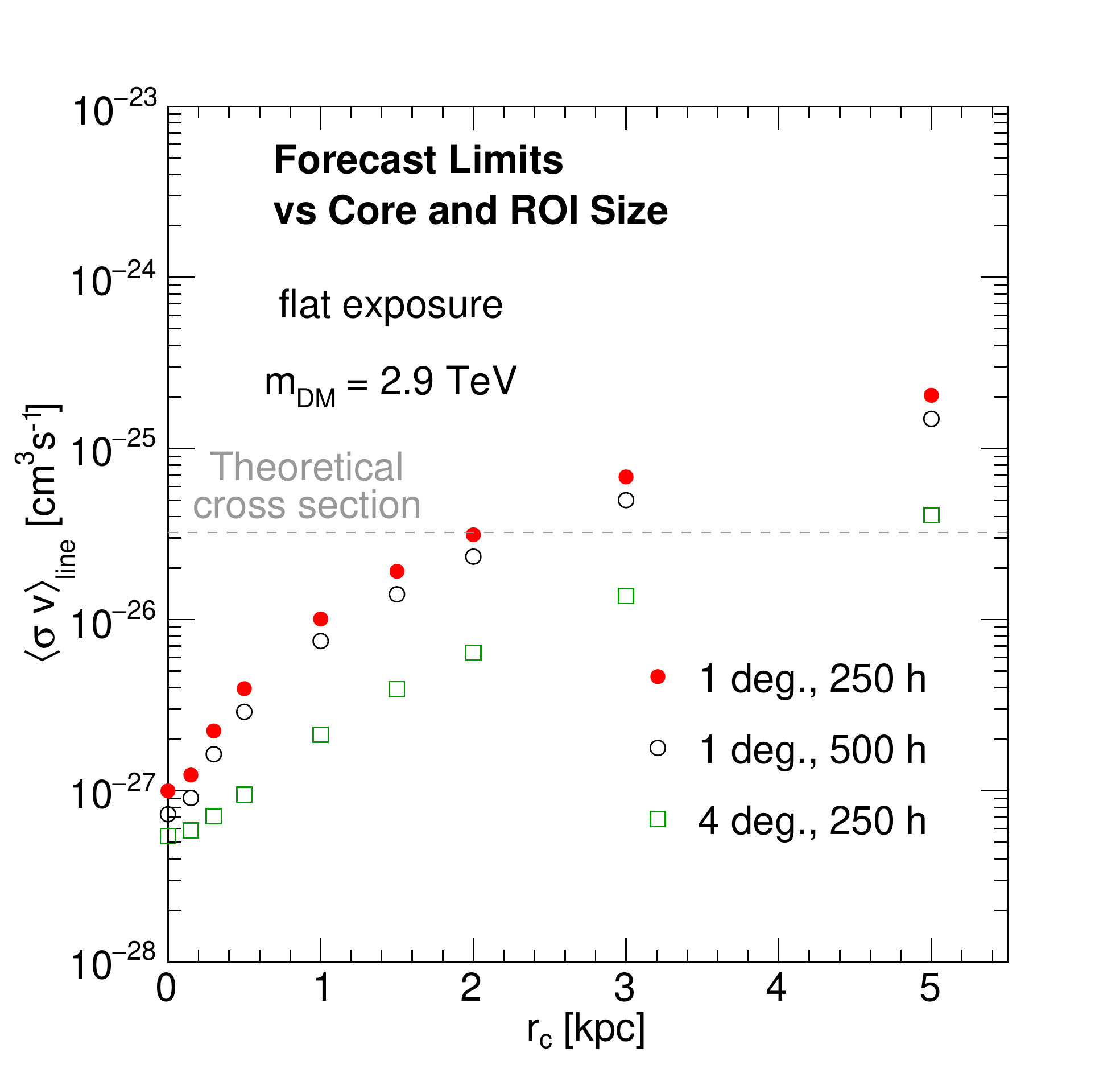}
\caption{{\it Left:} Mean expected upper limits at 95\% C.L. on the wino annihilation cross section $\langle \sigma\,v \rangle_\text{line}$ as a function of the DM mass $m_\DM$  assuming the full line + endpoint + continuum spectrum considering ROIs up to $4^\circ$, where the profile is varied as labeled in the legend.   A 250 hour homogeneous exposure is assumed over all the ROIs.  The NLL cross section for wino DM is shown in gray.
{\it Right:} Comparison of 95\% C.L. mean expected upper limits on $\langle \sigma\,v \rangle_\text{line}$ versus core radius size $r_{\rm c}$ for a 2.9 TeV wino using the IGS observation strategy with respect to the H.E.S.S.-I-like observation strategy. The limits assuming a 500 hour homogeneous exposure over the ROIs up to $1^\circ$ are also shown.}
\label{fig:prospects}
\end{center}
\end{figure*}

The right panel of~Fig.~\ref{fig:compare} shows the 95\% C.L. mean expected upper limits from the line signal, the line + endpoint spectrum, and the full spectrum given by the line + endpoint + continuum contributions. The limits for wino DM are driven by the line signal for TeV masses. The contribution of the endpoint signal compared to the line-only signal is significant and increases with the DM mass.  For DM masses of 2.3 TeV, 2.9 TeV and 9 TeV, the endpoint contribution improves the line-only limit by a factor 1.4, 1.5 and 2.1, respectively. The continuum signal 
is subdominant compared to the endpoint contribution, but its contribution increases with the DM mass. For a DM mass of 2.3 TeV, 2.9 TeV and 9 TeV, the continuum contribution improves the line plus endpoint limits by 8\%, 12\% and 27\%, respectively. 
 
Given our poor knowledge of the DM distribution in the GC region, it is useful to instead express the overall limits for pure wino DM (assuming it constitutes 100\% of the DM) as a limit on the total $J$-factor for the ROI. We show this $J$-factor limit as a function of the DM mass in Fig.~\ref{fig:Jcorelimit}. Due to the Sommerfeld enhancement, which yields resonances in the annihilation cross section at specific DM masses, very strong constraints on the $J$-factor are obtained at 2.3 and 9 TeV DM masses, respectively. 

The left panel of Fig.~\ref{fig:corelimit} shows the impact of a cored DM distribution in the GC on the 95\% C.L. mean expected limit on $\langle \sigma\,v \rangle_\text{line}$.
For cored profiles, the limits degrade by a factor up to 200 compared to the Einasto profile assuming core radii up to 5 kpc. For a 2.3 TeV DM mass, DM profiles with core radii lower than 5 kpc can be excluded. For DM mass of 9 TeV, DM profiles with core radii lower than 3 kpc can be excluded, as shown in the right panel of Fig.~\ref{fig:corelimit}. At the thermal DM mass of 2.9 TeV, the forecast limit on the core size is approximately 2 kpc.

The H.E.S.S. collaboration is pursuing an inner Galaxy survey (IGS) of the central several degrees of the GC region~\cite{Abdallah:2018qtu}. While the H.E.S.S.-I-like observations of the GC region were defined with pointing positions of the telescopes up to 1.5$^{\circ}$ from the Galactic plane, the observation strategy currently carried out utilizes pointing positions up to 3$^\circ$ in Galactic latitudes. We now compute projected expected limits considering all the ROIs up to 4$^{\circ}$ from the GC (IGS-like strategy) with a homogenous exposure over all the ROIs with observations assuming only phase-I telescopes for the gamma-ray event selection and reconstruction. We compute the signal and background counts in all the ROIs up to 4$^{\circ}$ following the procedure described in Sec.~\ref{sec:mockdata}. 
The left panel of Fig.~\ref{fig:prospects} shows the 95\% C.L. mean expected limit on $\langle \sigma\,v \rangle_\text{line}$ as a function of the wino mass for the DM profiles considered in this study. For cored DM  profiles the limits degrade only by a factor up to $\sim70$ compared to the Einasto profile assuming core radii up to 5 kpc. The dependence of the limits on the DM profile shape in the inner region of the Milky Way is less pronounced with the IGS observation strategy 
compared to the H.E.S.S-I-like one. 
The right panel of Fig.~\ref{fig:prospects} shows the 95\% C.L. mean expected limit on $\langle \sigma\,v \rangle_\text{line}$ as a function of the core radius size for a wino DM mass of 3 TeV. Using the IGS observation strategy, the limits improve significantly over the ones obtained from H.E.S.S.-I-like one. For the Einasto and 3 kpc core DM profile, the improvement is a factor of 1.8 and 5, respectively. The ratio of the IGS-like limits over the H.E.S.S.-I-like limits versus core radius size improves up to a core radius of $\sim 1$ kpc. Beyond this radius, the improvement follows the ratio between $J$-factors computed in 1$^{\circ}$  and 4$^{\circ}$ ROIs. For the thermal wino mass of 2.9 TeV, this translates to a projected core size limit of approximately 4.5 kpc. Doubling the overall exposure in the inner 1$^{\circ}$ would only slightly increase the limits. The limits improve by 37\% with respect to the H.E.S.S.-I-like limits independent of the DM profile core radius.

%%%%%%%%%%%%%%%%%%%%%%%%%%%%%%%%%%%%%%%
\section{Conclusions}
\label{sec:summary} 
%%%%%%%%%%%%%%%%%%%%%%%%%%%%%%%%%%%%%%%

We show the prospects for wino DM over a mass range from 1 TeV up to 70 TeV using VHE gamma-ray observations of the GC that rely on the most-up-to-date EFT computation of the annihilation spectrum of winos.  We build realistic mock data simulations of H.E.S.S.-I-like observations of the GC region and implement spectral and spatial analysis of the VHE emissions. We compute the sensitivity to wino DM using a binned likelihood test statistic ratio using the spectral and spatial information of signal and background. Various DM density distributions in the GC region are considered including DM density cores up to 5 kpc.

We show that ($i$) the line contribution to the wino annihilation spectrum drives the overall limits in the TeV mass range, ($ii$) the endpoint contribution significantly improves the sensitivity compared to the line-only signal with increasing importance for higher DM masses, and ($iii$) the continuum contribution is sub-dominant compared to the line and endpoint contributions  but becomes more relevant as the DM mass increases. 

The present sensitivity of H.E.S.S.-I-like observations is able to provide strong constraints on wino DM. We show for the case of winos constituting 100\% of the DM that strong constraints on the DM density in the central region of the Milky Way can be obtained using H.E.S.S.-I-like observations of the GC region. In particular, DM cores up to several kpc radii could be excluded for 2.3 and 9 TeV masses respectively, where the Sommerfeld effect strongly enhances the annihilation cross section through resonances.

We additionally provided a sensitivity projection for H.E.S.S.-I-like observations of the GC region using the IGS strategy. For the thermal wino mass of 2.9 TeV, we find a projected core size limit of approximately 4.5 kpc. This makes clear that future searches will provide a decisive test for the wino under reasonable assumptions for how the DM is distributed in the center of the Milky Way at kpc scales. 

%%%%%%%%%%%%%%%%%%%%%%%%%%%%%%%%%%%%%%%%%%%%%%%%%%%%%%%%%%%%%%%%%%%%%%%%%%%%%%%%
\begin{acknowledgments}

We are particularly grateful to M.~Solon for his collaboration on the leading log theory result.
MB is supported by the Office of High Energy Physics of the U.S. Department of Energy under Contract No. DE-SC-0000232627.
NLR and TRS are supported by the U.S. Department of Energy, under grant numbers DE-SC00012567 and DE-SC0013999.
NLR is further supported by the Miller Institute for Basic Research in Science.
IM is supported by the Office of High Energy Physics of the U.S. Department of Energy under Contract No. DE-AC02-05CH11231.
TC is supported by the U.S. Department of Energy, under grant numbers DE-SC0018191 and  DE-SC0011640.
IWS is supported by the Office of Nuclear Physics of the U.S. Department of Energy under the Grant No. DE-SCD011090 and by the Simons Foundation through the Investigator grant 327942.
VV is supported by the Office of Nuclear Physics of the U.S. Department of Energy under the Grant No.  Contract DE-AC52-06NA25396 and through the LANL LDRD Program.

\end{acknowledgments}
%%%%%%%%%%%%%%%%%%%%%%%%%%%%%%%%%%%%%%%%%%%%%%%%%%%%%%%%%%%%%%%%%%%%%%%%%%%%%%%%
\end{spacing}

\bibliography{bibl}
\bibliographystyle{utphys}

\end{document}